\documentclass[aps,prd,showpacs,nofootinbib,showkeys,superscriptaddress,preprintnumbers]{revtex4-1}
\usepackage{amsmath,amssymb,latexsym}
\usepackage{graphicx,epsfig,epstopdf}
\usepackage{bm,bbm}
\usepackage{braket,slashed,multirow}
\usepackage{natbib}
\usepackage{float}
\usepackage{color}
\usepackage[bookmarks=true,bookmarksopen=true,bookmarksnumbered=true,bookmarksopenlevel=3]{hyperref}

\newcommand{\be}{\begin{equation}}
\newcommand{\ee}{\end{equation}}
\newcommand{\bea}{\begin{eqnarray}}
\newcommand{\eea}{\end{eqnarray}}
\newcommand{\nn}{\nonumber}
\newcommand{\sh}[1]{#1\hspace{-6pt}/}
\def\Tr{{\rm Tr}}

\def\als{\alpha_{\rm s}} 
\def\MS{\overline{\rm MS}}
\def \nc {N_c}
\def \bp {\mathbf{p}}
\def \cf {C_F}

\begin{document}

\title{Heavy quarkonium suppression in a fireball}
\author{Nora Brambilla}
\affiliation{Physik Department, Technische Universit\"at M\"unchen, D-85748 Garching, Germany}
\affiliation{Institute for Advanced Study, Technische Universit\"at M\"unchen, Lichtenbergstrasse 2 a, D-85748 Garching, Germany}
\author{Miguel A.~Escobedo}
\affiliation{Department of Physics, P.O. Box 35, FI-40014 University of Jyv\"askyl\"a, Finland} 
\author{Joan Soto}
\affiliation{Departament de F\'\i sica Qu\`antica i Astrof\'\i sica and Institut de Ci\`encies del Cosmos, 
Universitat de Barcelona, Mart\'\i $\;$ i Franqu\`es 1, 08028 Barcelona, Catalonia, Spain}
\author{Antonio Vairo}
\affiliation{Physik Department, Technische Universit\"at M\"unchen, D-85748 Garching, Germany}

\date{\today}

\preprint{TUM-EFT 89/16}

\begin{abstract}
We perform a comprehensive study of the time evolution of heavy-quarkonium states in an expanding hot QCD medium by implementing effective field theory techniques 
in the framework of open quantum systems. The formalism incorporates quarkonium production and its subsequent evolution in the fireball including quarkonium dissociation and recombination. 
We consider a fireball with a local temperature that is much smaller than the inverse size of the quarkonium and much larger than its binding energy.
The calculation is performed at an accuracy that is leading-order in the heavy-quark density expansion and next-to-leading order in the multipole expansion. 
Within this accuracy, for a smooth variation of the temperature and large times, the evolution equation can be written as a Lindblad equation.
We solve the Lindblad equation numerically both for a weakly-coupled quark-gluon plasma and a strongly-coupled medium.
As an application, we compute the nuclear modification factor for the $\Upsilon(1S)$ and $\Upsilon(2S)$ states.
We also consider the case of static quarks, which can be solved analytically.
Our study fulfills three essential conditions: it conserves the total number of heavy quarks,
it accounts for the non-Abelian nature of QCD and it avoids classical approximations.
\end{abstract}

\maketitle

\section{Dilepton emission from quarkonium}
\label{sec:dilepton}
The main way in which heavy quarkonium is detected in heavy-ion collisions is through its decay into a lepton pair. 
Electromagnetic interactions are slow compared with the strong interactions that drive the dynamics of the fireball; therefore, the bulk of these decays will happen after freeze-out. 
An observation that supports this understanding is that the positions of the
peaks in the dilepton emission spectrum are the same in pp and AA collisions~\cite{Chatrchyan:2012lxa}.

The decay rate into leptons in thermal equilibrium was computed in~\cite{McLerran:1984ay}.
Here we generalize that result to the case of a medium that is not in thermal equilibrium. 
The Hamiltonian of the system can be written as $H=H_{\rm QCD}+H_{\rm EW}$,  
where $H_{\rm QCD}$ is the QCD Hamiltonian, and $H_{\rm EW}$ is the part of the Hamiltonian that includes leptons and the electroweak interaction 
(for the following use only the electromagnetic part of the electroweak interaction is relevant). 
At some early time $t_0$ we have a system that only contains quarks and gluons. 
It is described by a density matrix $\rho(t_0)=\sum_{nm}\rho_{nm}(t_0)|n\rangle \langle m|$, where $|n\rangle$ are eigenstates of $H_{\rm QCD}$.
At a much later time the system is made by an arbitrary number of quarks and gluons and a lepton pair, $l^+l^-$. 
Hence, up to higher orders in the electromagnetic coupling, $\alpha$, 
the state describing such a system is the product of a QCD state, $|n\rangle$, and a lepton-pair state,  $|l^+l^-\rangle$: $|n,l^+l^-\rangle \equiv |n\rangle |l^+l^-\rangle$. 
Naming $k_1$ and $k_2$ the momenta of the two outgoing leptons, the differential emission rate is
\begin{equation}
d\mathcal{R}=\sum_{n,m,j} \rho_{nm}(t_0)\langle m |j,l^+(k_1)l^-(k_2)\rangle\langle j,l^+(k_1)l^-(k_2) | n\rangle\,d^3k_1d^3k_2\,.
\end{equation}
At leading order in $\alpha$, we can write 
\begin{equation}
\langle j,{l^+}^s(k_1){l^-}^{s'}(k_2)|n\rangle = e^2 e_Q\int\,d^4x_1\,d^4x_2
\phantom{b}_{\rm EW}\langle 0|a^s_{k_1}b^{s'}_{k_2}\,\bar{l}(x_1)\sh{A}(x_1) l(x_1)\, A_\nu(x_2)|0\rangle_{\rm EW} \;\langle j|J^\nu(x_2)|n\rangle\,,
\end{equation}
where $|0\rangle_{\rm EW}$ is the vacuum of the electroweak theory, 
$a_{k_1}^s$ and $b_{k_2}^{s'}$ are the lepton and antilepton annihilation operators, 
$l$ is the lepton field, $A_\mu$ the photon field, $J^\mu$ the quark electromagnetic current, 
and $e_Q$ the fraction of electron charge $e$ carried by the heavy quark.
Summing over the polarizations, we obtain 
\begin{equation}
d\mathcal{R}=-\frac{e^4e_Q^2L_{\mu\nu}(k_1,k_2)}{|{\bf k_1}||{\bf k_2}|(k_1+k_2)^4}\int\,d^4x_1\,d^4x_2\,e^{-i(k_1+k_2)\cdot(x_1-x_2)}
\,\Tr\left\{\rho(t_0)\, J^\mu(x_1)J^\nu(x_2)\right\}\,d^3k_1d^3k_2\,,
\label{dR2}
\end{equation}
where
\begin{equation}
L_{\mu\nu}(k_1,k_2) = k_1 \cdot k_2 g_{\mu\nu} - k_{1\mu}k_{2\nu} - k_{2\mu}k_{1\nu}\,.
\end{equation}
For $\rho=e^{-H_{\rm QCD}/T}$, where $T$ is a temperature, we recover the results of~\cite{McLerran:1984ay}.

In general, the medium formed in a heavy-ion collision will be out of thermal equilibrium 
and characterized by some correlation lengths that change with time.
In this work, in order to keep close contact with and take advantage of existing studies of the medium in thermal equilibrium, 
we will often assume that the medium evolves in time in a quasistatic fashion.
This means that the medium is locally in thermal equilibrium and that at each time we can define a temperature $T$.
This assumption allows us to take over results obtained in thermal equilibrium, with the only difference being that now the temperature depends on time. 
Our explicit model for the time dependence of the temperature will be discussed in Sec.~\ref{sec:Bj}.
A quantitative characterization of a temperature varying slowly (quasistatically) with time in the context of heavy quarkonium dissociation will be given at the end of Sec.~\ref{sec:evolution}. 
Nevertheless, it is worth stressing that the evolution equations that we will derive in Sec.~\ref{sec:evolution} do not rely on the quasistatic approximation 
and that in some of the following discussions, including the rest of this introduction, we may simply understand $T$ as the inverse of a correlation length characterizing the system.

In order to write more explicitly Eq.~\eqref{dR2}, it is convenient to take advantage of the 
heavy-quark mass, $M$,  being much larger than the typical momentum of the particles in the fireball, which is proportional to $T$\footnote{
  The scale induced by the thermal bath is really $\pi T$, however here and in the following we will just write $T$ for brevity.}:
\begin{equation}
M \gg T\,.
\end{equation}
One can therefore use non-relativistic QCD (NRQCD) to describe the heavy-quark dynamics~\cite{Caswell:1985ui,Bodwin:1994jh}. 
In NRQCD, heavy quarks are represented by a Pauli field $\psi$ that annihilates the heavy quark 
and a Pauli field $\chi$ that creates the heavy antiquark. 
Up to corrections of order $\als(M)$ and $T/M$, the NRQCD electromagnetic current reads 
\bea
J^0(x) &=& \psi^\dagger(x)\psi(x)-\chi^\dagger(x)\chi(x)\,, 
\\
J^i(x) &=& e^{i2M v\cdot x}\psi^\dagger(x)\sigma^i\chi(x)-e^{-i2M v\cdot x}\chi^\dagger(x)\sigma^i\psi(x)\,.
\eea
In the quarkonium rest frame, it is $v=(1,{\bf 0})$.
$J^0$ does not contribute to the emission of the lepton pair with an invariant energy around $2M$. 
This contribution comes only from $J^i$. 
The Pauli matrix in $J^i$ projects onto the subspace of quarkonia with spin $1$.
In terms of the NRQCD heavy-quark fields, the emission rate can be written as
\begin{equation}
d\mathcal{R}=\frac{e^4e_Q^2L_{ij}(k_1,k_2)}{|{\bf k_1}||{\bf k_2}|(k_1+k_2)^4}\int\,d^4x_1\,d^4x_2\,e^{-i(k_1+k_2-2 M v)\cdot(x_1-x_2)}\,
\Tr\left\{\rho(t_0)\,\psi^\dagger(x_1)\sigma^i\chi(x_1)\chi^\dagger(x_2)\sigma^j\psi(x_2)\right\}\,d^3k_1d^3k_2\,.
\end{equation}

The right-hand side of the previous equation will depend on different energy scales, some coming from the thermal plasma, 
which depend on the temperature $T$, and some from the non-relativistic nature of the heavy-quark-antiquark bound state.
These last ones are the typical heavy-quark-antiquark distance, $a_0$, and the typical quark-antiquark binding energy, $E$.
They respect the non-relativistic hierarchy: 
\begin{equation}
M \gg \frac{1}{a_0} \gg E\,.
\end{equation}
The physics of heavy-quarkonium in the fireball is going to depend on the relation between these two sets of scales. 
We will assume in the following that 
\begin{equation}
\frac{1}{a_0} \gg T\,.
\label{hiecoulombic}
\end{equation}

Under the condition \eqref{hiecoulombic}, we can use potential NRQCD (pNRQCD), which provides a valid description of the 
quarkonium at energy scales below $1/a_0$~\cite{Pineda:1997bj,Brambilla:1999xf,Brambilla:2004jw,Brambilla:2008cx,Escobedo:2008sy}. 
In this Effective Field Theory (EFT), the heavy-quark-antiquark system can be described in terms of a color-singlet field $S$ and a color-octet field $O$ 
instead of the fields $\psi$ and $\chi$ of NRQCD. 
At leading order, 
\begin{equation}
\chi_\alpha^\dagger(t,{\bf x}_1)\psi_\beta(t,{\bf x}_2) \to S_{\alpha\beta}(t,{\bf r}, {\bf R}) \,,
\end{equation}
where ${\bf r} = {\bf x}_1 -{\bf x}_2$; ${\bf R} = ({\bf x}_1 + {\bf x}_2)/2$; and $\alpha$, $\beta$ are spin indices.
The projection of the color-singlet field on $S$-wave states may be decomposed into a spin zero component
($\delta_{\alpha\beta} \Tr\{S\}/\sqrt{2}$) and a spin one component ($\sigma^\ell_{\alpha\beta} \Tr\{S\sigma^\ell\}/\sqrt{2}$).
Only the spin one component contributes to the $S$-wave emission rate ($R=(t,{\bf R})$, $R'=(t',{\bf R}')$):
\begin{equation}
d\mathcal{R}|_{\hbox{\tiny S-wave}} = \frac{2e^4e_Q^2L_{ii}(k_1,k_2)}{3|{\bf k_1}||{\bf k_2}|(k_1+k_2)^4}\int d^4R\,d^4R'\, e^{-i(k_1+k_2-2 M v)\cdot(R-R')}\,
\Tr\left\{\rho(t_0)\, S^{\ell\,\dagger}(t,0,{\bf R})S^\ell(t',0,{\bf R}')\right\}\,d^3k_1d^3k_2\,,
\end{equation}
where $S^\ell \equiv  \Tr\{S\sigma^\ell\}/\sqrt{2}$.
The result reflects the expectation that the emission of dileptons should be proportional to the number of singlet states with spin $1$.

The integration of the temporal components $t$ and $t'$ may be split into  
\begin{equation}
\int_{t_0}^{t_F} dt \,(\cdots)  + \int_{t_F}^{\infty} dt \, (\cdots) \,,
\end{equation}
where $t_F$ is the time of freeze-out. The argument that most of dilepton production from quarkonium happens in the vacuum amounts to having 
\begin{equation}
\int_{t_F}^{\infty} dt \, (\cdots)  \gg \int_{t_0}^{t_F} dt \, (\cdots) \,.
\label{eq:appro}
\end{equation}
After freeze-out, the system evolves as it would in the vacuum. Hence for $t$, $t'$ larger than $t_F$, we have 
\begin{eqnarray}
&&\Tr\left\{ \rho(t_0)\, S^{\ell\,\dagger}(t,{\bf 0},{\bf R})S^\ell(t',{\bf 0},{\bf R}') \right\}\equiv \langle {\bf 0},{\bf R}'|\rho_s(t';t)|{\bf 0},{\bf R}\rangle= 
\nonumber \\
&&\sum_{nm} e^{i E^*_m (t-t_F)} e^{-iE_n(t'-t_F)} 
\phi^*_m({\bf 0}) \phi_n({\bf 0})
\int\frac{d^3P }{(2\pi)^3}e^{i\frac{{\bf P}^2}{4M}(t-t_F) - i{\bf P}\cdot{\bf R}} \,
\int\frac{d^3P'}{(2\pi)^3}e^{-i\frac{{\bf P}^{\prime 2}}{4 M}(t'-t_F) + i{\bf P}'\cdot{\bf R}'}
\langle n,{\bf P}'|\rho_s(t_F;t_F)|m,{\bf P}\rangle\,, 
\nn\\
\label{dTF}
\end{eqnarray}
where $\phi_n$ ($\phi_m$) is the wave function of the $n$th ($m$th) state, $E_n$ ($E_m$) is the corresponding energy (including the decay width as an imaginary phase), 
and ${\bf P}$ (${\bf P'}$) the center-of-mass momentum; $\rho_s(t';t)$ will be defined in full generality later on (see~\eqref{rhosdef}) and will be the central object of our study.
Therefore under the approximation \eqref{eq:appro} and in the quarkonium rest frame, we have 
\begin{equation}
d\mathcal{R}|_{\hbox{\tiny S-wave}} = \frac{2e^4e_Q^2L_{ii}(k_1,k_2)}{3|{\bf k_1}||{\bf k_2}|q^4}
\sum_{nm}\frac{\phi^*_m({\bf 0}) \phi_n({\bf 0})\, \langle n,{\bf q}|\rho_s(t_F;t_F)|m,{\bf q}\rangle}
{[q_0 - 2 M - E_m^* - {\bf q}^2/(4M)] [q_0 - 2M - E_n   - {\bf q}^2/(4M)]}\,d^3k_1d^3k_2\,,
\end{equation}
where we have defined $q=k_1+k_2$.

The above equation tells us that the emission rate of dileptons with the energy of a given quarkonium state $n$ ($q^0 = 2M + {\rm Re}(E_n)+{\bf q}^2/4M$)
will be proportional to the projection of $\rho_s(t_F;t_F)$ into that state at freeze-out time. 
The problem reduces to find $\rho_s(t_F;t_F)$ given some initial conditions at $t=0$, $\rho_s(0;0)$. 
Within this framework, the yield of quarkonium $nS$ states in $AA$ collisions normalized with respect to the yield in $pp$ collision is given by   
\begin{equation}
R_{AA}(nS) = \frac{\langle n,{\bf q}|\rho_s(t_F;t_F)|n,{\bf q}\rangle}{\langle n,{\bf q}|\rho_s(0;0)|n,{\bf q}\rangle}\,.
\label{eq:raa}
\end{equation} 
Note that the ratio above depends in principle on the center-of-mass momentum of the quarkonium state ${\bf q}$. We will eventually neglect this dependence and write $|nS\rangle$ for $|n,{\bf q}\rangle$.

The paper is structured in the following way. In Sec.~\ref{sec:Bj} we briefly describe the evolution of the medium in a simple model.
In Sec.~\ref{sec:tevol} we derive the general evolution equations for the heavy-quark-antiquark densities in the framework of pNRQCD.
This section contains the main theoretical results of the paper. 
In Sec.~\ref{sec:weak} we write the Lindblad equation and solve it for a quarkonium in a weakly-coupled plasma that fulfills the hierarchy $1/a_0 \gg T \gg E \gg m_D$, 
where $m_D$ is the Debye mass (for details see Appendix~\ref{E>m_D}) , 
while in Sec.~\ref{sec:strong} we do the same for a quarkonium in a strongly-coupled plasma that fulfills the hierarchy $1/a_0 \gg T \sim m_D \gg E$ 
(the weakly-coupled case $1/a_0 \gg T \gg m_D \gg E$ is also addressed as a particular case, see Appendix~\ref{m_D>E}). 
Numerical results are presented for the bottomonium states $\Upsilon(1S)$ and $\Upsilon(2S)$.
The analytical solution of the Lindblad equation in the case of static sources is derived in Appendix~\ref{app1}.
Finally, in Sec.~\ref{sec:con} we draw some conclusions and discuss possible developments.
A concise version of the evolution equations and their solution in the case of a strongly-coupled plasma has been presented in~\cite{Brambilla:2016wgg}.

\section{Time evolution of the thermal medium}
\label{sec:Bj}
We will consider a medium that is infinite, homogeneous and isotropic in space but that changes with time. 
These approximations are appropriate for very large nuclei in central collisions. 
For the description of these kinds of systems we can use Bjorken's evolution~\cite{Bjorken:1982qr}. 
These assumptions about the medium could be relaxed by describing the medium using relativistic hydrodynamics (see~\cite{Ollitrault:2008zz} for a review); 
however, this is not the largest source of uncertainty in our calculation (higher-order corrections, uncertainties related to the hierarchy of scales, 
and the assumed Coulombic nature of some excited quarkonium states appear to have a larger impact).

According to~\cite{Bjorken:1982qr} the effective temperature of the system evolves in time following 
\begin{equation}
T=T_0\left(\frac{t_0}{t}\right)^{v_s^2}\,,
\label{Tbj}
\end{equation}
where $T_0$ and $t_0$ are, respectively, the initial temperature and initial (proper) time, 
and $v_s$ is the velocity of sound in the medium. 
In a deconfined plasma at a very high temperature, $v_s^2= 1/3$. 
As values of $T_0$ and $t_0$ for central collisions at the LHC, we use $T_0=475$~MeV and $t_0=0.6$~fm. 
These values are taken from~\cite{Alberico:2013bza}.

We will study collisions with different centralities. 
As we are assuming the plasma to be homogeneous and isotropic, 
the only effect that a difference in centrality will produce will be to change
the initial value of the energy density and hence $T_0$. 
For this we assume the initial energy density to be proportional to $T_0^4$ 
(this is consistent with the sound velocity that we are using). 
Moreover, we assume the initial energy density to depend on the number of participants. 
From (2.6) of~\cite{Kolb:2000sd} we obtain
\begin{equation}
T_0(b)=T_0(b=0)\left(\frac{T_A(b/2,0)\left[1-\left(1-\frac{\sigma T_A(b/2,0)}{A}\right)^A\right]}
{T_A(0,0)\left[1-\left(1-\frac{\sigma T_A(0,0)}{A}\right)^A\right]}\right)^{1/4}\,,
\label{T0b}
\end{equation}
where $b$ is the impact parameter; $A$ the number of nucleons; $\sigma$ the nucleon scattering cross-section, 
which, according to~\cite{Chatrchyan:2011sx}, is taken as $\sigma=64\pm 5$~mb $=164\pm 13$~GeV$^{-2}$ and
\begin{equation}
T_A(x,y)=\int_{-\infty}^\infty dz\,\rho_A(x,y,z)\,,
\end{equation}
where $\rho_A$ is the nucleon density distribution in the nucleus. 
Following~\cite{Chatrchyan:2011sx}, we approximate $\rho_A$ by a Woods--Saxon profile 
(assuming that the nucleon density is proportional to the charge density) taken from~\cite{devries}:
\begin{equation}
\rho_A(x,y,z) \approx \frac{\rho_0}{1+e^{\frac{r-c}{\xi}}}\,,
\end{equation}
where $r^2 = x^2 + y^2 +z^2$; $c=6.62\pm 0.06$ fm; $\xi=0.546\pm 0.01$ fm; 
and $\rho_0$ is fixed by $\displaystyle \int d^3r \,\rho_A = A$, which, in the case of lead at the LHC, is 207.

\begin{table}[ht]
\begin{center}
\begin{tabular}{|c|c|c|}
\hline
Centrality (\%) & $\langle b\rangle$ (fm) & $T_0$ (MeV) \\ 
\hline
$0-20$ & $4.76$ & $466$ \\
$0-10$ & $3.4$ & $471$ \\
$10-20$ & $6.0$ & $461$ \\
$20-90$ & $11.6$ & $360$ \\
$20-30$ & $7.8$ & $449$ \\
$30-40$ & $9.35$ & $433$ \\
$40-50$ & $10.6$ & $412$ \\
$30-50$ & $9.9$ & $425$ \\
$50-70$ & $12.2$ & $366$ \\
$50-100$ & $13.6$ & $304$ \\
\hline
\end{tabular}
\end{center}
\caption{Initial temperature of the fireball for different centrality bins and its mean impact parameter. 
When we say $0-10\%$ centrality, we mean that from all the collisions we select the $10\%$ that are most central.
\label{tab:central}}
\end{table}

\begin{figure}
\includegraphics[scale=0.4]{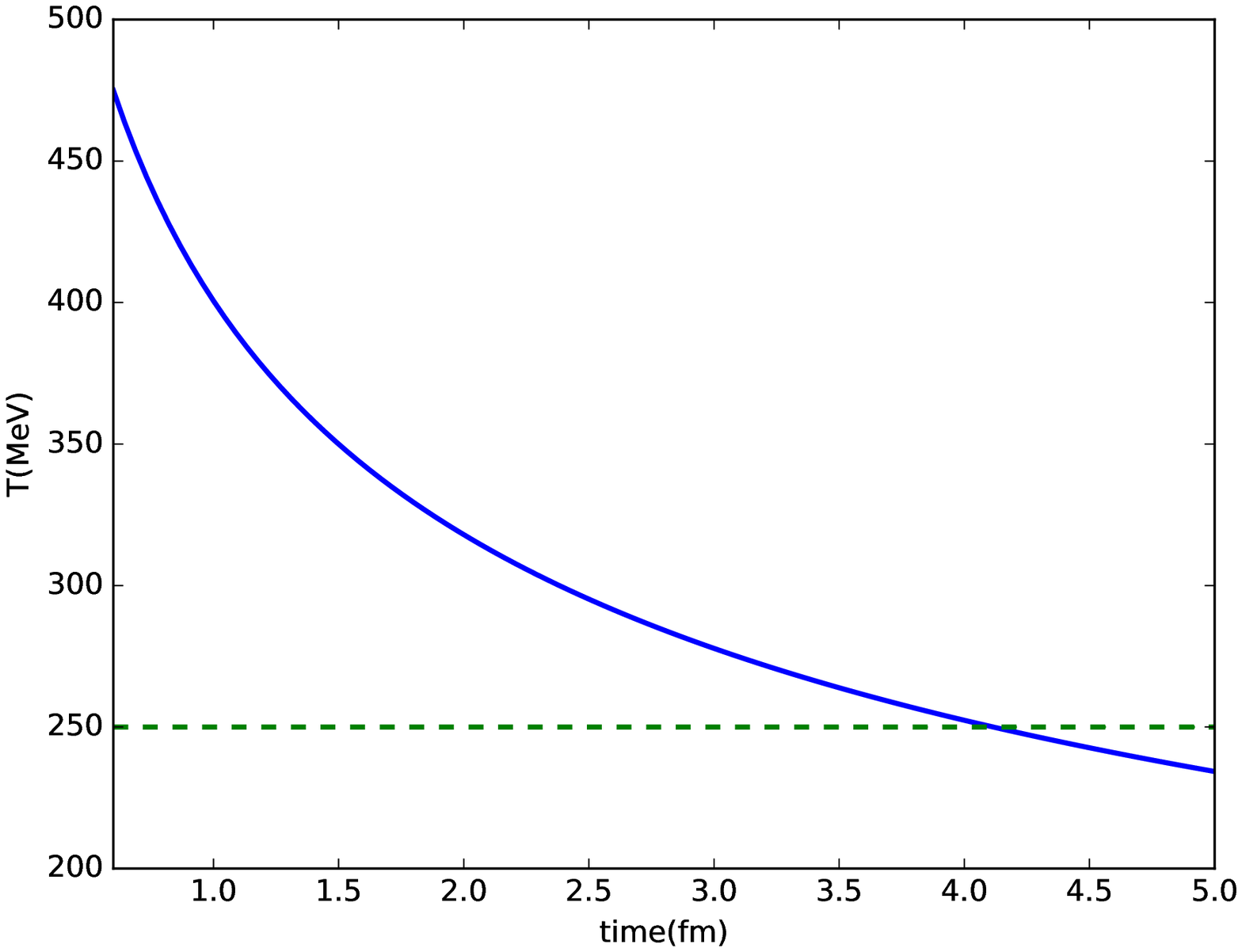}\includegraphics[scale=0.4]{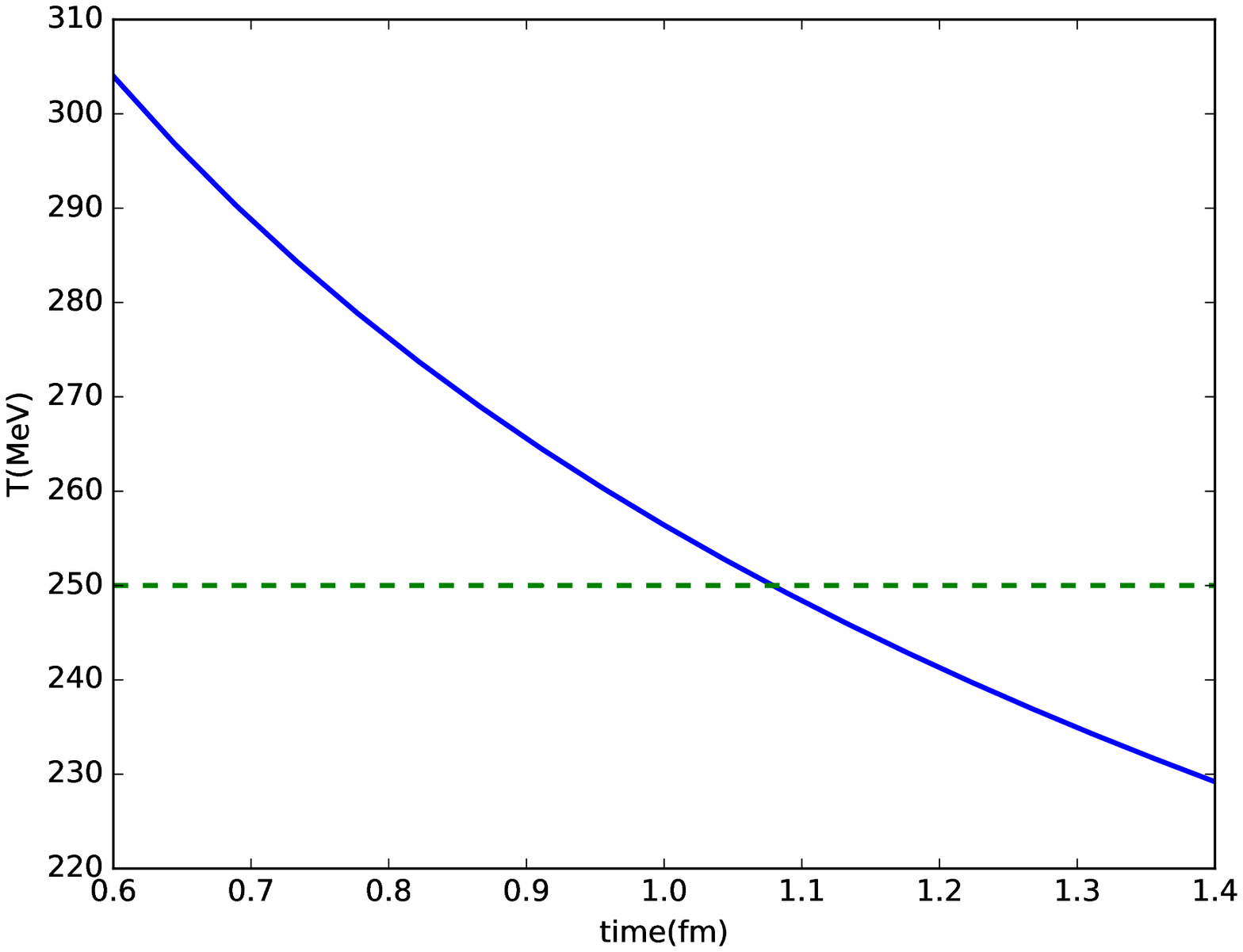}
\caption{Time evolution of the temperature according to \eqref{Tbj} 
for the most central (left) and the most peripheral (right) collisions of Table~\ref{tab:central}.
\label{fig:Ttime}
}
\end{figure}

Experimentally, the results are given in terms of centrality bins. 
When possible ~\cite{Chatrchyan:2011sx}, we characterize each centrality bin used in CMS by its mean impact parameter.
There are some centrality bins whose mean impact parameter is not given in ~\cite{Chatrchyan:2011sx}. In these cases we perform the computation ourselves, following ~\cite{Miller:2007ri}.
In all calculations in the rest of the paper we will describe the different centralities by the different $T_0$ given in Table~\ref{tab:central}.
For all the centrality bins given in this table except $20-90 \%$, the value of $T_0$ corresponds to $T_0(\langle b\rangle)$ (which is similar to $\langle T_0(b)\rangle$).
In the case of $20-90 \%$ centrality, the value given is $\langle T_0(b)\rangle$ due to the fact that the bin is so large that it is not well represented by the average impact parameter.
In Fig.~\ref{fig:Ttime} we show the time evolution of the temperature according to \eqref{Tbj} for the most and least central collisions of Table~\ref{tab:central}.
The evolution starts at $t_0=0.6$~fm and ends at about 4~fm for the most central collisions and at about 1.1~fm for the most peripheral ones, 
when the fireball has cooled down to a temperature of about 250~MeV, which is the smallest temperature, 
but still larger than the crossover temperature to the quark-gluon plasma, where we expect the framework of our calculation to be safely realized.
Indeed, we will need $T\gg E$ to be always fulfilled, and, in the weakly-coupled plasma case, we will also need $T$ large enough to trust our perturbative calculations.
In the strong coupling case, the last limitation does not apply, 
but in practice the lack of lattice data below $1.5 T_c$ for the relevant non-perturbative quantities forces us to stick to the above-mentioned temperature.
We recall that the crossover temperature to the quark-gluon plasma, $T_c$, is expected to be at about 150~MeV~\cite{Borsanyi:2010bp,Bazavov:2014pvz,Bazavov:2016uvm}.
Henceforth, we will compute the quarkonium nuclear modification factors $R_{AA}$ at the earlier time when a temperature of about 250~MeV is reached.
Clearly the last piece of the evolution from 250~MeV to the freeze-out temperature is missing.
Nevertheless, we expect it to have a modest impact on our results.

\section{Time evolution of the quarkonium density matrix for $1/a_0 \gg T$}
\label{sec:tevol}
The typical LHC temperature of the fireball goes from around 475~MeV to the freeze-out temperature. 
The energy scale induced by the temperature is $\pi T$, which is at most about 1.5~GeV.
Since the fireball expands very fast at the beginning of the thermal evolution, it will reach very soon lower energies. 
The inverse of the Bohr radius, $1/a_0$,  of the $\Upsilon(1S)$ is in the vacuum about 1.3 GeV.
The conclusion is that, for most of the evolution of the fireball (except perhaps for a very short time at the beginning), 
the condition \eqref{hiecoulombic} is fulfilled at least by the bottomonium ground state.
Moreover $1/a_0 \approx 1.3$~GeV$\gg \Lambda_{\rm QCD}$ implies that the bound state is Coulombic, i.e., described by a Coulomb potential.

\subsection{pNRQCD for an open quantum system}
Under the condition \eqref{hiecoulombic} and the assumption that the bound state is Coulombic, $1/a_0 \gg \Lambda_{\rm QCD}$, 
we can describe the quarkonium evolution in the fireball using pNRQCD. 
The Lagrangian of pNRQCD at next-to-leading order in the multipole expansion is~\cite{Pineda:1997bj,Brambilla:1999xf,Brambilla:2004jw} 
\bea
\mathcal{L}_{\rm pNRQCD} \!&=&\! \int d^3r \, 
\Tr\left[S^\dagger\left(i\partial_0-h_s\right)S+O^\dagger\left(iD_0-h_o\right)O\right]
+ \Tr\left[ O^\dagger{\bf r}\cdot g{\bf E}S+S^\dagger{\bf r}\cdot g{\bf E}O 
+\frac{1}{2} (O^\dagger{\bf r}\cdot g{\bf E}O+O^\dagger O{\bf r}\cdot g{\bf E}) \right]
\nonumber \\
\!&& \!
+\mathcal{L}_{\rm light} \,,
\label{eq:Lagrangian_pNRQCD}
\eea
where $r$ is the distance between the heavy quark and the antiquark,
and ${\rm S} = S\,\mathbbm{1}_c/\sqrt{N_c}$ and ${\rm O}=\sqrt{2}O^aT^a$ stand for the heavy quark-antiquark fields in a color-singlet 
and color-octet configuration, respectively.
The operator $h_s = {\bf p}^2/M + V_s$ is the color-singlet Hamiltonian.  
The potential $V_s$ is the color-singlet potential, which at leading order reads $V_s = - C_F \als(1/a_0)/r$; 
$C_F=(N_c^2-1)/(2N_c)= 4/3$ is the Casimir of the fundamental representation and $N_c=3$ is the number of colors.
The operator $h_o = {\bf p}^2/M + V_o$ is the color-octet Hamiltonian. 
The potential $V_o$ is the color-octet potential, which at leading order reads $V_o =  \als(1/a_0)/(2N_cr)$. 
We have made manifest that the strong coupling in the potentials is evaluated at a scale that is of the order of the inverse Bohr radius.
We have set equal to 1 the Wilson coefficients of the dipole operators 
(corrections are suppressed by powers of $\als$ and are beyond our aimed leading-order accuracy).
The term $\mathcal{L}_{\rm light}$ is the QCD Lagrangian with light quarks.

In \eqref{eq:Lagrangian_pNRQCD} there is a covariant derivative acting on the octet field $O$.
This can be eliminated by means of suitable field redefinitions: $O=\Omega O'\Omega^\dagger$ and ${\bf E}=\Omega {\bf E'}\Omega^\dagger$. 
The only effect of them is to change $D_0O$ into $\partial_0O'$ in \eqref{eq:Lagrangian_pNRQCD} and to rename the 
fields $O$ and ${\bf E}$ into $O'$ and ${\bf E}'$. 
The field $\Omega$ can be chosen to be a Wilson line going from $-\infty$ to $t$: 
$\displaystyle \Omega = \exp\left[ -ig \int_{-\infty}^t ds \, A_0(s,{\bf R}) \right]$.
In the following, we will adopt these field redefinitions and we will understand the octet field and the chromoelectric field as the redefined ones. 
We drop, however,  the apex $'$ to simplify the notation.

As discussed in Sec.~\ref{sec:dilepton}, we want to compute $\Tr\left\{ \rho(t_0) \, S^{\ell\,\dagger}(t,{\bf r},{\bf 0})S^\ell(t,{\bf r}',{\bf 0})\right\}$. 
The experimental fact that the number of bottom quarks found in heavy-ion collisions is much smaller than that of lighter quarks 
implies that $\Tr\{\rho(t_0)\}$ is dominated by contributions to the density matrix, $\rho$, coming from states made of light quarks and gluons.
This is clearly the case in thermal equilibrium,
where contributions coming from the heavy quarks are suppressed by a factor $e^{-M/T}$. 
A consequence of this is that 
$\Tr\left\{\rho(t_0)S^{\ell\, \dagger}(t,{\bf r},{\bf 0})S^\ell(t',{\bf r}',{\bf 0})\right\}\ll\Tr\left\{\rho(t_0)S^\ell(t',{\bf r}',{\bf 0})S^{\ell\, \dagger}(t,{\bf r},{\bf 0})\right\}$ 
and similarly for the octet field. 

We may look at the heavy quarks as an open quantum system that interacts with the (slowly) evolving medium of the fireball made of light quarks and gluons.
The computation can be done in the close-time-path formalism
(see~\cite{lebellac} for the thermal equilibrium version and~\cite{Berges:2004yj} for the non-equilibrium case).
The formalism consists in rewriting field correlators by allowing the time to evolve 
from a path that goes from $-\infty$ to $\infty$ and then from $\infty-i\epsilon$ to $-\infty-i\epsilon$. 
The fields are then ordered along the path. 
To make the ordering manifest, we call the fields on the upper branch type $1$ and those on the lower branch type~$2$, and we identify them by a corresponding index. 
Hence, we write 
\begin{eqnarray}
\Tr\{ \rho(t_0) \, S^\dagger(t, {\bf r}, {\bf R}) S(t', {\bf r}', {\bf R}')\} 
&=&  \langle S_1(t',{\bf r}',{\bf R}') S_2^\dagger(t,{\bf r},{\bf R})\rangle 
\equiv \langle {\bf r}',{\bf R}'|\rho_s(t';t)|{\bf r},{\bf R}\rangle,
\label{rhosdef}
\\
\Tr\{ \rho(t_0) \, O^{a\dagger}(t, {\bf r}, {\bf R}) O^b(t', {\bf r}', {\bf R}')\} 
&=&  \langle O_1^b(t',{\bf r}',{\bf R}') O_2^{a\dagger}(t,{\bf r},{\bf R})\rangle 
\equiv 
\langle {\bf r}',{\bf R}'|\rho_o(t';t)|{\bf r},{\bf R}\rangle \frac{\delta^{ab}}{N_c^2-1},
\label{rhoodef}
\end{eqnarray}
where in each equation the last equality defines the color-singlet and the color-octet density respectively.
Since we do not have a preferred direction in color space, the octet density is taken as diagonal in color.

We further assume that the heavy quarks are comoving with the medium. 
Under this simplifying assumption, the evolution does not depend on ${\bf R}$ and we drop it from the arguments of the densities.
To simplify the notation, we have also dropped the dependence on the heavy quark-antiquark distance;
nevertheless, the densities should be understood as operators that depend on it.

We emphasize that the situation that we consider here is different from the one studied in~\cite{Brambilla:2008cx,Escobedo:2008sy}, 
where the  system formed by the plasma plus the heavy-quark states was assumed to be in thermal equilibrium.
Under that condition the density of heavy-quark states is exponentially suppressed and, 
when computing heavy-quark correlators, we only need to include heavy-quark fields living on the upper branch of the closed-time path. 
This is not the case here, where the number of heavy-quark states is not the equilibrium distribution.
As a technical remark, we further observe that the 12 propagator does not select a specific time ordering, 
while the 11 and 22 propagators describing heavy-quark fields living on the upper and lower branch, respectively,
select instead the forward and backward propagation: they are proportional to $\theta(t'-t)$ and $\theta(t-t')$, respectively.

\subsection{Evolution equations}
\label{sec:evolution}
We assume the following simplified model for the evolution of the quarkonium in the medium. \\

{\it (a)} From $t=0$ (which could also be taken as a time in the infinite past, $t = -\infty$) until $t=t_0$, the heavy quarks evolve as in the vacuum. 
Therefore up to corrections of relative order $a_0^2 E^3$, which are negligible with respect to thermal corrections as long as $T \gg E$,
the color-singlet and color-octet densities evolve as
\begin{eqnarray}
\rho_{s}(t';t) &=& e^{-ih_s t'} \, \rho_{s}(0;0) \, e^{ih_s t}\,,
\label{vacuumrhos}\\
\rho_{o}(t';t) &=& e^{-ih_o t'} \, \rho_{o}(0;0) \, e^{ih_o t}\,.
\label{vacuumrhoo}
\end{eqnarray}
The singlet and octet Hamiltonians have to be understood as operators in the relative-distance space like the densities.
The initial conditions, $ \rho_{s,o}(0;0)$, of the densities will be discussed in Sec.~\ref{sec:initial}. \\

{\it (b)} At $t=t_0$, suddenly the medium appears and the heavy-quarks start interacting with it.
We model the medium as a fireball that follows Bjorken's time evolution (see Sec.~\ref{sec:Bj}).
Since the number of heavy quarks in the medium is relatively small, 
we organize the computation as an expansion in the heavy-quark densities, $\rho_s$ and $\rho_o$.
The $12$ propagators are proportional to these densities. 
We compute them by keeping only Feynman diagrams with a single $12$ propagator of heavy quarks, 
which amounts to considering only terms that are linear in the heavy-quark densities. 
Diagrams that contain two $12$ correlators are quadratic in the densities of heavy quarks, and so on. 
Since we compute only diagrams that are linear in the heavy-quark densities, 
we consistently ignore the density dependence in the 11 and 22 propagators too. 

\begin{figure}[ht]
\includegraphics[scale=0.4]{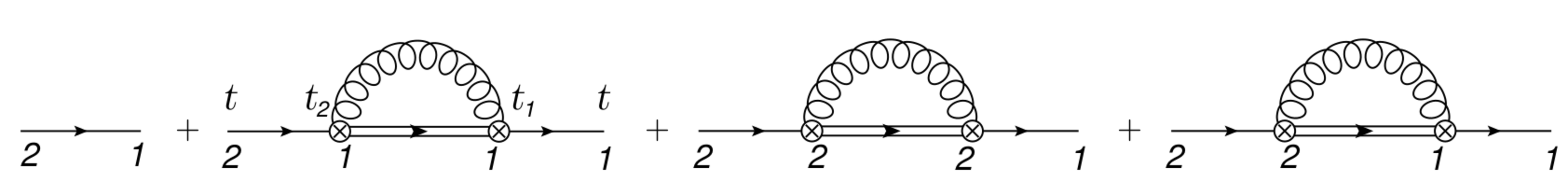}
\caption{Diagrams contributing at order $r^2$ to $\rho_s$. 
A single line stands for a singlet propagator, a double line for an octet propagator, and a curly line for gluons. 
The vertices (circle with a cross) are the chromoelectric dipole vertices of the pNRQCD Lagrangian \eqref{eq:Lagrangian_pNRQCD}.
The numbers $1$ or $2$ near the vertices mean insertions of fields from the upper or lower branches of the closed-time path, respectively. 
In the second diagram we also write explicitly the time variables of the propagators according to Eq.~\eqref{feyndiag2}.
\label{fig:noresum}
}
\end{figure}

The initial conditions $\rho_{s,o}(t_0;t_0)$ are determined from the evolution starting at $t=0$ and ending at $t=t_0$ computed in~{\it (a)}.
We calculate now corrections to $\rho_s(t;t)$ for $t>t_0$ at order $r^2$. 
The relevant diagrams are shown in Fig.~\ref{fig:noresum}. 
From the pNRQCD Lagrangian it follows that at zeroth-order in the multipole expansion, $\rho_{s}(t;t)$ is just given by the tree-level diagram
\be
\hspace{-4.7cm}
\includegraphics[scale=0.5]{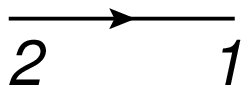} \put(4,8){$= e^{-ih_s (t-t_0)}\rho_s(t_0;t_0)e^{ih_s (t-t_0)}.$}
\label{feyndiag1}
\ee

\noindent 
If the initial and final times are different, then the tree-level 12 propagator reads $e^{-ih_s (t-t_0)}\rho_s(t_0;t_0)e^{ih_s (t'-t_0)}$.
Such propagators enter in the following one-loop diagrams.

The second diagram in Fig.~\ref{fig:noresum} reads
\be
\hspace{-8cm}
\includegraphics[scale=0.5]{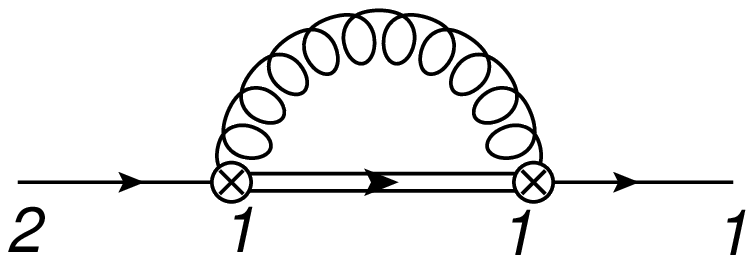}\put(4,9){$\displaystyle  = - \int_{t_0}^t dt_1 \, e^{-ih_s(t-t_1)} \, \Sigma_s(t_1) \, e^{-ih_s(t_1-t_0)} \, \rho_s(t_0;t_0) \, e^{i h_s(t-t_0)}\,,$}
\label{feyndiag2}
\ee

\noindent
with 
\begin{equation}
\Sigma_s(t) = \frac{g^2}{2N_c}\int_{t_0}^{t} dt_2 \,  r^i \, e^{-ih_o (t-t_2)} \, r^j \, e^{ih_s (t-t_2)} \, \langle E^{a,i}(t,{\bf 0})E^{a,j}(t_2,{\bf 0})\rangle \,.
\label{eq:sigma}
\end{equation}
The expression  $\langle E^{a,i}(t,{\bf 0})E^{a,j}(t_2,{\bf 0})\rangle$, like similar expressions below, 
stands for the in-medium light degrees of freedom average of the correlator of two chromoelectric fields. 
We recall that the fields in the previous expression are the redefined ones and hence gauge invariant by themselves.
In this work, we assume that the medium is isotropic in space and color, and locally and instantaneously in thermal equilibrium at a temperature~$T$. 
The third diagram in Fig.~\ref{fig:noresum} is the complex conjugate of this one:
\be
\hspace{-8cm}
\includegraphics[scale=0.5]{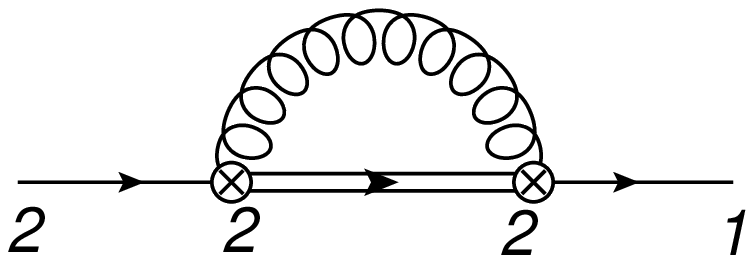}\put(4,9){
$\displaystyle  =-\int_{t_0}^t dt_1\, e^{-ih_s(t-t_0)} \, \rho_s(t_0;t_0) \, e^{ih_s(t_1-t_0)}\,\Sigma_s^\dagger(t_1)\,e^{ih_s(t-t_1)}\,.$}
\label{feyndiag3}
\ee

\noindent
Finally the last diagram in Fig.~\ref{fig:noresum} reads
\be
\hspace{-8cm}
\includegraphics[scale=0.5]{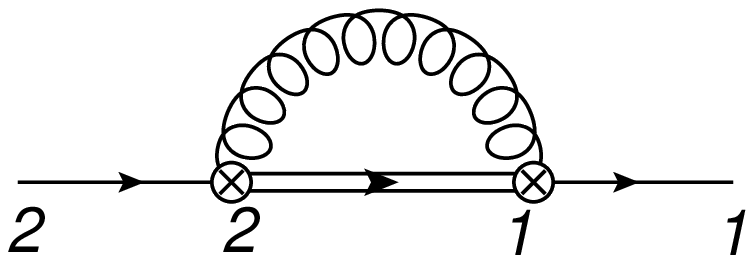}\put(4,9){
$\displaystyle  = \int_{t_0}^t dt_1\,  e^{-ih_s(t-t_1)} \, \Xi_{so}(\rho_o(t_0;t_0),t_1)\, e^{ih_s(t-t_1)}\,,$}
\label{feyndiag4}
\ee

\noindent
with 
\begin{equation}
\Xi_{so}(\rho_o(t_0;t_0),t) =  \frac{g^2}{2N_c(N_c^2-1)} \int_{t_0}^{t} dt_2 \, \left[r^i \, e^{-ih_o(t-t_0)}\, \rho_o(t_0;t_0)\, e^{ih_o(t_2-t_0)} \, r^j \, e^{ih_s (t-t_2)} \, 
\langle E^{a,j}(t_2,{\bf 0})E^{a,i}(t,{\bf 0})\rangle 
+ \hbox{H.c.}
\right],
\label{eq:Xi}
\end{equation}
where H.c. stands for Hermitian conjugate.

Taking into account all the diagrams in Fig.~\ref{fig:noresum} and deriving both sides with respect to $t$, we get the evolution equation for $\rho_s$:
\begin{eqnarray}
\frac{d\rho_s(t;t)}{dt} &=& -i[h_s,\rho_s(t;t)] - \Sigma_s(t)e^{-ih_s(t-t_0)}\rho_s(t_0;t_0)e^{ih_s(t-t_0)} - e^{-ih_s(t-t_0)}\rho_s(t_0;t_0)e^{ih_s(t-t_0)}\Sigma_s^\dagger(t) 
\nonumber\\
&& + \Xi_{so}(\rho_o(t_0;t_0),t)\,.
\label{eq:evolsinglet}
\end{eqnarray}

A similar computation leads to the evolution equation for $\rho_o$:
\begin{eqnarray}
\frac{d\rho_o(t;t)}{dt} &=& -i[h_o,\rho_o(t;t)]-\Sigma_o(t)e^{-ih_o(t-t_0)}\rho_o(t_0;t_0)e^{ih_o(t-t_0)}-e^{-ih_o(t-t_0)}\rho_o(t_0;t_0)e^{ih_o(t-t_0)}\Sigma_o^\dagger(t)
\nonumber\\
&&+\Xi_{os}(\rho_s(t_0;t_0),t)+\Xi_{oo}(\rho_o(t_0;t_0),t)\,,
\label{eq:evoloctet}
\end{eqnarray}
with
\begin{eqnarray}
\Sigma_o(t) &=& 
\frac{g^2}{2 N_c(N_c^2-1)}\int_{t_0}^t dt_2 \, r^i \, \left[ e^{-ih_s (t-t_2)}  + \frac{N_c^2-4}{2} e^{-ih_o (t-t_2)} \right] \, r^j \, e^{ih_o (t-t_2)} 
\, \langle E^{a,i}(t,{\bf 0})E^{a,j}(t_2,{\bf 0})\rangle \,,
\label{eq:sigmao}
\end{eqnarray}
\begin{equation}
\Xi_{os}(\rho_s(t_0,t_0),t) = 
\frac{g^2}{2N_c} \int_{t_0}^t dt_2  \, \left[r^i \, e^{-ih_s(t-t_0)}\, \rho_s(t_0;t_0) \, e^{ih_s(t_2-t_0)} \, r^j \, e^{ih_o (t-t_2)} \, \langle E^{a,j}(t_2,{\bf 0})E^{a,i}(t,{\bf 0})\rangle
+ \hbox{H.c.}
\right],
\label{eq:Xi1}
\end{equation}
\begin{equation}
\Xi_{oo}(\rho_o(t_0;t_0),t) =
\frac{g^2 (N_c^2-4)}{4 N_c (N_c^2-1)} 
\int_{t_0}^t dt_2  \, \left[r^i\, e^{-ih_o(t-t_0)} \, \rho_o(t_0;t_0) \, e^{ih_o(t_2-t_0)} \, r^j \, e^{ih_o (t-t_2)} \, \langle E^{a,j}(t_2,{\bf 0})E^{a,i}(t,{\bf 0})\rangle
+ \hbox{H.c.}
\right].
\label{eq:Xi2}
\end{equation}

\begin{figure}[ht]
\includegraphics[scale=0.35]{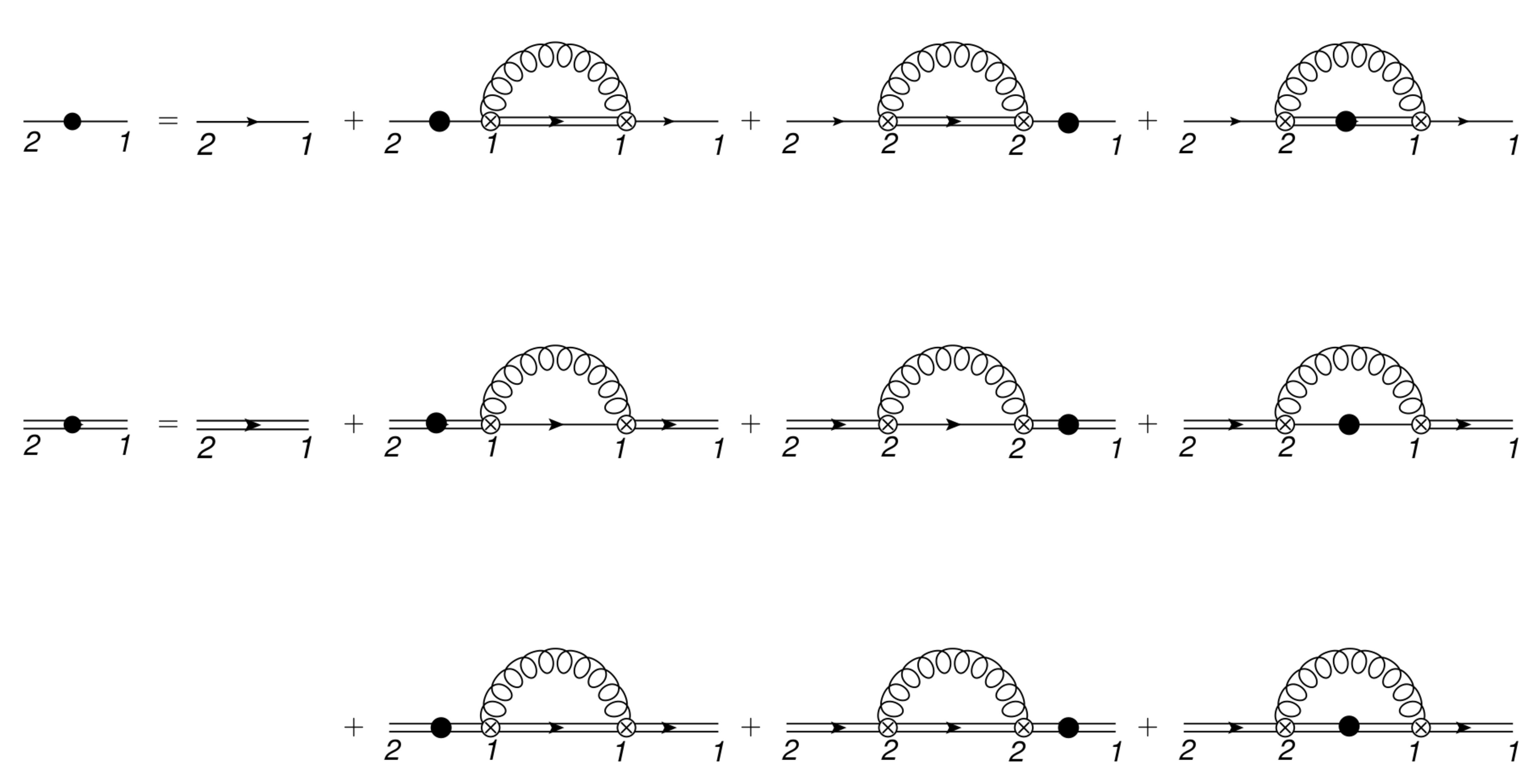}
\caption{Coupled Schwinger--Dyson equations for the singlet and octet 12 propagators.
\label{fig:resum}
}
\end{figure}

Both Eqs. \eqref{eq:evolsinglet} and \eqref{eq:evoloctet} are such that their right-hand sides depend on $e^{-ih_{s,o}(t-t_0)}\,\rho_{s,o}(t_0;t_0)\,e^{ih_{s,o}(t-t_0)}$.
This combination is $\rho_{s,o}(t;t)$ at tree level.
The question arises whether one should solve the equations above as they stand or if one should substitute $e^{-ih_{s,o}(t-t_0)}\,\rho_{s,o}(t_0;t_0)\,e^{ih_{s,o}(t-t_0)}$ by $\rho_{s,o}(t;t)$.
First, one notices that substituting $e^{-ih_{s,o}(t-t_0)}\,\rho_{s,o}(t_0;t_0)\,e^{ih_{s,o}(t-t_0)}$ by $\rho_{s,o}(t;t)$ in Eqs. \eqref{eq:evolsinglet} and \eqref{eq:evoloctet}
induces new terms at order $r^4$ only, hence this substitution is consistent with the accuracy of those equations, which is of order $r^2$.
Furthermore, promoting  $e^{-ih_{s,o}(t-t_0)}\,\rho_{s,o}(t_0;t_0)\,e^{ih_{s,o}(t-t_0)}$ to  $\rho_{s,o}(t;t)$ in the right-hand sides of Eqs. \eqref{eq:evolsinglet} and \eqref{eq:evoloctet}
makes the equations Markovian and, in particular, leads to the standard evolution equations for the density matrix in the case of unitary evolution ($\Xi_{s,o}=0$, $\Sigma_{s,o}^\dagger=-\Sigma_{s,o}$).
Finally, it has the advantage of providing a set of equations that hold for arbitrarily large times. 
One should notice in fact that the validity of the expansion in Fig.~\ref{fig:noresum} relies upon the restriction $(t-t_0) \Sigma, (t-t_0) \Xi, ... \ll 1$.
This restriction is lifted by the promotion. 
After the promotion of $e^{-ih_{s,o}(t-t_0)}\,\rho_{s,o}(t_0;t_0)\,e^{ih_{s,o}(t-t_0)}$ to  $\rho_{s,o}(t;t)$,
the system of equations for the singlet- and octet-density evolutions takes the form
\begin{eqnarray}
\frac{d\rho_s(t;t)}{dt} &=& -i[h_s,\rho_s(t;t)] - \Sigma_s(t)\rho_s(t;t) - \rho_s(t;t)\Sigma_s^\dagger(t)
+\Xi_{so}(\rho_o(t;t),t)\,,
\label{eq:ev} \\
\frac{d\rho_o(t;t)}{dt} &=& -i[h_o,\rho_o(t;t)] - \Sigma_o(t)\rho_o(t;t) - \rho_o(t;t)\Sigma_o^\dagger(t)
+\Xi_{os}(\rho_s(t;t),t)+\Xi_{oo}(\rho_o(t;t),t)\,,
\label{eq:ev_octet}
\end{eqnarray}
where $\Xi_{so}$, $\Xi_{os}$, and $\Xi_{oo}$ are as in Eqs. \eqref{eq:Xi}, \eqref{eq:Xi1} and \eqref{eq:Xi2} with the replacement $e^{-ih_{s,o}(t-t_0)}\rho_{s,o}(t_0;t_0)e^{ih_{s,o}(t-t_0)}\to\rho_{s,o}(t;t)$.
The interactions of the density matrices with the medium are characterized by only three independent operators, as we make explicit in Appendix~\ref{linear}. 
The equations above are equivalent to the Schwinger--Dyson equations represented in Fig.~\ref{fig:resum}.\footnote{If $\rho_{s,o}(t;t)$ is written as $\rho_{s,o}(t;t)=e^{-ih_{s,o}(t-t_0)}$ $\rho^I_{s,o}(t;t)$ $e^{ih_{s,o}(t-t_0)}$, then $\rho^I_{s,o}$ is the density matrix in the interaction picture. 
At tree level $\rho^I_{s,o}(t;t) = \rho_{s,o}(t_0;t_0)$, but loop corrections modify this relation and make the time evolution of $\rho_{s,o}(t;t)$ sensitive to two different time scales. 
One will come from the energy exponentials $e^{\pm ih_{s,o}(t-t_0)}$ and will be of the order of the inverse of the binding energy, $1/E$,  
whereas the other will come from $\rho^I_{s,o}(t;t)$ and will be of the order of $1/(\als a_0^2\Lambda^3)$, due to the pNRQCD power counting.  
$\Lambda$ is generically the next relevant scale in the system ($T$ or $E$ or a combination of them), hence $1/a_0 \gg \Lambda$.
Since $1/(\als a_0^2\Lambda^3)\gg 1/E$, the evolution of $\rho^I_{s,o}$ may be considered slow.} 

The interpretation of the functions appearing in (\ref{eq:ev})  and (\ref{eq:ev_octet}) is clear. 
The self energies $\Sigma_s$ and $\Sigma_o$ provide the in-medium induced mass shifts, $\delta m_{s,o}$, 
and widths, $\Gamma_{s,o}$, for the color-singlet and color-octet heavy quark-antiquark systems, respectively:
\begin{eqnarray}
-i\Sigma_{s,o}(t) + i\Sigma_{s,o}^\dagger(t)  &=& 2 \, {\rm Re}\,(-i\Sigma_{s,o}(t)) = 2 \delta m_{s,o}(t),
\label{deltam}\\
\Sigma_{s,o}(t) + \Sigma_{s,o}^\dagger(t)  &=& -2 \, {\rm Im}\,(-i\Sigma_{s,o}(t)) = \Gamma_{s,o}(t).
\label{gammaself}
\end{eqnarray}
The function $\Xi_{so}$ accounts for the production of singlets through the decay of octets, 
while the functions $\Xi_{os}$ and $\Xi_{oo}$ account for the production of octets through the decays of singlets and octets respectively. 
There are two octet production mechanisms that can eventually be traced back to the two octet chromoelectric dipole vertices in the pNRQCD Lagrangian \eqref{eq:Lagrangian_pNRQCD}. 

Finally, we note that the conservation of the trace of the sum of the densities, which amounts to the conservation of the number of heavy quarks, 
requires the functions $\Sigma_s$, $\Sigma_o$, $\Xi_{so}$, $\Xi_{os}$ and $\Xi_{oo}$ to satisfy 
\begin{eqnarray}
  \Tr \left\{\rho_s(t;t) \left(\Sigma_s(t) + \Sigma_s^\dagger(t)\right)\right\} &=& \Tr \left\{\Xi_{os}(\rho_s(t;t),t) \right\}\,, 
\label{conssin}\\
\Tr \left\{\rho_o(t;t) \left(\Sigma_o(t) + \Sigma_o^\dagger(t)\right)\right\} &=& \Tr \left\{ \Xi_{so}(\rho_o(t;t),t) + \Xi_{oo}(\rho_o(t;t),t) \right\}\,.
\label{consoct}
\end{eqnarray}
The above equations, relating the singlet and octet decay widths to the corresponding production matrix elements, represent nothing else than the optical theorem in the problem at hand.
They are fulfilled by the expressions in \eqref{eq:sigma}, \eqref{eq:Xi}  and \eqref{eq:sigmao}-\eqref{eq:Xi2}, which becomes apparent if they are written like in Appendix \ref{linear}.

In the following we will assume $t-t_0$ to be larger than any other time scale that appears in the problem.
This is indeed so for any time of the order of magnitude of the freeze-out time.
It also amounts to assuming that the time during which the subsystem is observed is much
larger than the time scale of any correlation between the subsystem and the environment.
Under this assumption, we will approximate 
\begin{equation}
\int_{t_0}^tdt_2  \, f(t_2) \approx \int_{0}^\infty ds \, f(t-s)  \,.
\label{approx1}
\end{equation}
Furthermore, we will approximate
\begin{equation}
\langle E^{a,i}(t,{\bf 0})E^{a,j}(t -s,{\bf 0})\rangle 
\approx \langle E^{a,i}(s,{\bf 0})E^{a,j}(0,{\bf 0})\rangle
\approx \langle E^{a,i}(0,{\bf 0})E^{a,j}(-s,{\bf 0})\rangle
\,.
\label{approx2}
\end{equation}
This is an exact equality in vacuum and in thermal equilibrium.  
In the out of equilibrium case that we are considering here, the equality is broken by the time dependence of the temperature. 
Nevertheless, if the evolution of the temperature is quasistatic, which is our case at large times
(cf. \eqref{Tbj} $1/T \times dT/dt \sim 1/t \ll E$), the time dependence of the temperature may be neglected at leading order.
One should point out that this approximation may be problematic at early times.

\subsection{Lindblad equation}
\label{sec:lindblad}
Equations \eqref{eq:ev} and \eqref{eq:ev_octet} form a system of equations that can be solved if the properties of the medium 
and the initial conditions for $\rho_s$ and $\rho_o$ are known.  This set of equations is the main result of this work. 
Although it is possible to solve these equations numerically for given initial conditions, it is indeed very challenging and computationally expensive.
For this reason we have chosen here to focus on cases where these equations can be simplified to a Lindblad equation~\cite{Lindblad:1975ef,Gorini:1975nb}. 
The Lindblad equation is well known in the fields of quantum optics and quantum information. 
It was studied in relation with quarkonium in~\cite{Akamatsu:2014qsa}. 
From a mathematical point of view, the Lindblad equation follows from requiring the time evolution of the density matrix of the open quantum system to be Markovian, 
to preserve the trace, the equation to be linear in the density, and the corresponding linear operator to be a completely positive map.
It has the following form:
\begin{equation}
\frac{d\rho}{dt}=-i[H,\rho]+\sum_n \left( C_n\,\rho \,C_n^\dagger-\frac{1}{2}\{C_n^\dagger C_n,\rho\} \right)\,,
\label{eq:Lindblad}
\end{equation}
where $H$ is a Hermitian operator and the operators $C_n$ are called collapse operators.
In our case, $\rho$ is the matrix 
\begin{equation}
\rho=\left(\begin{array}{c c}
\rho_s & 0 \\
0 & \rho_o\end{array}\right)\,.
\end{equation}

\subsection{Expansion in spherical harmonics}
\label{sec:spherical}
We solve the Lindblad equation using numerical libraries available in the literature~\cite{Johansson20121760,Johansson20131234} and putting the system on a lattice. 
However, this lattice is three dimensional, making the number of entries for operators in the Lindblad equation still prohibitively large for our present means. 
As a way to deal with this practical difficulty, we do an expansion in terms of spherical harmonics. We define
\begin{equation}
\rho^{lm;l'm'}=\int\,d\Omega(\hat{r})\,d\Omega(\hat{r}') \, Y^{lm}(\hat{r}) \, \rho \, {Y^{l'm'}}^*(\hat{r}')\,.
\end{equation}
Since there is no preferred direction in space, during the entire evolution only the components with $l=l'$ and $m=m'$ are non-zero. 
Moreover, by the same argument, all polarizations are equally possible; therefore, all the information can be encoded in 
\begin{equation}
\rho^l \equiv \sum_m\rho^{lm;lm}\,.
\end{equation} 
The normalization of $\rho^l$ has been chosen to ensure that $\sum_l \rho^l = \Tr\{\rho\}$.
The crucial approximation that simplifies the numerics of the calculation is to consider only $l=0$ and $l=1$. 
We have checked that the results change very little if we also include $l=2$.
The reason is that we are interested in the suppression factor of $S$-wave states. 
Via a chromoelectric dipole transition these states can decay to or be generated by $P$-wave states only.
Hence, they are affected by states with $l$ larger than $1$ only indirectly.
Under this approximation the density matrix can be written as
\begin{equation}
\rho=\left(\begin{array}{cccc}
\rho^0_s & 0 & 0 & 0\\
0 & \rho^1_s & 0 & 0\\
0 & 0 & \rho^0_o & 0 \\
0 & 0 & 0 & \rho^1_o \end{array}\right)\,.
\end{equation}
The corresponding Hamiltonian is 
\begin{equation}
h^l_{s,o}=-\frac{1}{M}\left( \frac{\partial^2}{\partial r^2}  + \frac{2}{r}\frac{\partial}{\partial r} \right) + V_{s,o} + \frac{l(l+1)}{M r^2}\,.
\label{hamiltonian_so}
\end{equation}

\subsection{Initial conditions for $\rho_s$ and $\rho_o$}
\label{sec:initial}
The computation of the production cross-section of quarkonium in pp and pA and its extrapolation to AA collisions is a non-trivial problem
and an active topic of research (see~\cite{Brambilla:2010cs} and references therein).
An additional difficulty is that results for production are usually written as expectation values for the different quarkonium states,  
while what we need is the reduced density matrix, which contains more information 
(for example the relative phase between the different states). 
Moreover, the thermalization procedure from the collision time until the hydrodynamic regime and 
the way in which quarkonium is affected by the medium between these two times are still largely unknown.

Facing these difficulties, we choose to make a naive assumption about the form of the initial conditions.
We impose that the relation between $\rho_s$ and $\rho_o$ at the initial time be controlled by just one parameter $\delta$. 
The fact that the creation of heavy quarks requires high energies tells us that 
singlets and octets will be formed in a configuration similar to a Dirac delta, which implies an $S$-wave state;
this assumption can also be found in~\cite{CasalderreySolana:2012av}. 
Because our evolution equations are linear in the densities and we are interested in $R_{AA}$, which is a ratio, 
we do not need to care about the absolute size of $\rho_s$ and $\rho_o$ but only about their relative size. 
The production cross-section of singlets in $S$-wave states was computed at leading order in~\cite{Baier:1983va} and that of octets in~\cite{Cho:1995ce}. 
Since the production of singlets is $\als$ suppressed compared to that of octets, we use as an initial condition that at collision time ($t=0$) 
\begin{eqnarray}
\rho_s(0;0) &=& N\, |{\bf 0}\rangle \, \langle {\bf 0}|,
\label{rhosinitial}\\
\rho_o(0;0) &=&  \rho_s(0;0) \, \frac{\delta}{\als(M)},
\label{rhooinitial}
\end{eqnarray}
where $|{\bf 0}\rangle$ is an eigenstate of ${\bf r}$ with eigenvalue ${\bf r}={\bf 0}$. 
The normalization $N$ is fixed by $\Tr\{\rho_s\} + \Tr\{\rho_o\} = 1$.
In the following we will perform the computations using different values of $\delta$ ($1$, $0.1$, and $10$).

Between the collision of the two ions at $t=0$ and the beginning of the hydrodynamic evolution at $t=0.6$~fm 
we assume that quarkonium evolves as if it were in vacuum. We use the bottom mass $M=4.8$~GeV. 
We obtain the Bohr radius for the 1$S$ bottomonium state from the condition $1/a_0 =  M C_F \als(1/a_0)/2$.
Using the $\MS$ scheme and one-loop running with $\Lambda_{\MS}=250$~MeV, we get $1/a_0 = 1.334$~GeV. 
The pNRQCD singlet and octet Hamiltonians are given after \eqref{eq:Lagrangian_pNRQCD} and for each angular momentum component in \eqref{hamiltonian_so}. 
For the singlet and octet potentials $V_s$ and $V_o$, we use the expressions given below (\ref{eq:Lagrangian_pNRQCD})
in accordance with the tree-level matching of pNRQCD with NRQCD. 
To compute the evolution of $\rho_s$ and $\rho_o$ (in vacuum and in the medium) we use a lattice of size $40a_0$ and spacing $a_0/10$. 
For the numerical computation we make use of the library Qutip~\cite{Johansson20121760,Johansson20131234}.

\section{Quarkonium in a weakly-coupled plasma: $1/a_0 \gg T \gg E \gg m_D$}
\label{sec:weak}
In this section, we derive the Lindblad equation and solve it for a weakly-coupled quark-gluon plasma in a particular thermodynamical regime.
A plasma is weakly coupled if $T \gg m_D$, where $m_D$ is the Debye mass.
Because in perturbation theory $m_D \sim g(T)\,T$, the condition is indeed fulfilled if the coupling $g(T)$ is small.
If there is no dynamical scale in between $T$ and $m_D$ the evolution equations are a particular case
of the strongly-coupled case, $T \sim m_D$, that we will address in more generality in the next section.
Explicit expressions can be found in Appendix~\ref{m_D>E}.
 
Here we consider the particular thermodynamical regime where the binding energy is in between $T$ and $m_D$.
This means that we consider a system that fulfills the hierarchy of scales: $1/a_0 \gg T \gg E \gg m_D,\,\Lambda_{\rm QCD}$.
In this situation and for thermal equilibrium, the modifications to the heavy quarkonium dynamics have been studied in detail~\cite{Brambilla:2010vq}. 
Both the energy levels and the decay widths get thermal corrections.
The thermal corrections to the width are mostly due to gluodissociation~\cite{Brambilla:2011sg}
(the dissociation of a quarkonium through the scattering with a real time-like gluon from the medium),
which at temperatures such that $E \gg m_D$ is a dissociation mechanism more important than the dissociation by inelastic parton scattering~\cite{Brambilla:2013dpa} 
(the dissociation of the quarkonium through scattering with partons in the medium).
In~\cite{Vairo:2010bm} it has been argued that this hierarchy of scales might be realized by the $\Upsilon(1S)$ state produced in heavy-ion collisions at the LHC.

Whereas it is clear that the real and imaginary parts of the singlet potential computed in~\cite{Brambilla:2010vq} must be related to $\Sigma_s$, 
it may not be obvious that they can be taken to be the same. 
In Appendix~\ref{E>m_D} we show that up to a redefinition of the density matrices, this is indeed the case.
An additional motivation for this redefinition is to obtain a system that can be solved using a Lindblad equation. 
We show in Appendix~\ref{linear} that, in general, \eqref{eq:ev} and \eqref{eq:ev_octet} are not positive, and hence cannot be written in the Lindblad form.
However, if we consider a medium that changes from thermal equilibrium in a quasistatic way,
and an evolution time large enough so that $\Sigma_s$ only depends on the medium through the temperature (see the discussion at the end of Sec.~\ref{sec:evolution}), 
then \eqref{eq:ev} and \eqref{eq:ev_octet} simplify and can be brought to the Lindblad form by the same redefinitions that led to the thermal equilibrium results of~\cite{Brambilla:2010vq}.
Taking over those results by just allowing a slow change of temperature with time,  
we see that they read at leading order
\begin{eqnarray}
{\rm Re}\,(-i\Sigma_s) &=& \frac{\pi}{9}N_cC_F \, \als(\mu_T)\, \als(1/a_0) \, T^2\, r + \frac{2\pi}{3M}C_F \, \als(\mu_T)\, T^2\,,
\label{eq:vs}\\
{\rm Im}\,(-i\Sigma_s) &=& - \frac{N_c^2C_F \, \als(\mu_E) \, \als^2(1/a_0)\, T}{6} - \frac{4 N_c C_F \, \als(\mu_E) \, \als(1/a_0) \, T}{3M r} - \frac{8C_F \, \als(\mu_E) \, T \, {\bf p}^2}{3M^2}\,,
\label{weak1}
\end{eqnarray}
where we have distinguished between the strong coupling running with the temperature scale, $\mu_T=\pi T$; the one running with the scale of the energy, $\mu_E \gtrsim -E_0 = 1/(M a_0^2)$; 
and the one coming from the Coulomb potential running with the inverse of the Bohr radius.

The thermal modifications to the octet potential were not calculated in~\cite{Brambilla:2010vq}. 
The calculation is very similar to the one for the singlet. 
We present it in Appendix~\ref{octet} for completeness. 
The result reads at leading order
\begin{eqnarray}
{\rm Re}\,(-i\Sigma_o) &=& -\frac{\pi}{18} \, \als(\mu_T) \, \als(1/a_0) \, T^2 \, r
+ \frac{\pi}{3M N_c} \, \als(\mu_T) \, T^2+ \frac{\pi}{6M} \,\frac{N_c^2-4}{N_c}\, \als(\mu_T) \, T^2\,,
\label{eq:vo}\\
{\rm Im}\,(-i\Sigma_o) &=& - \frac{N_c\, \als(\mu_E)\, \als^2(1/a_0)\, T}{12} 
+ \frac{2 \, \als(\mu_E) \, \als(1/a_0) \, T}{3M r}
- \frac{4\, \als(\mu_E) \, T \, {\bf p}^2}{3N_c M^2} - \frac{2(N_c^2-4) \, \als(\mu_E)\, T\, {\bf p}^2}{3N_c M^2}\,.
\label{weak2}
\end{eqnarray}

The different $\Xi$'s must be computed directly from \eqref{eq:Xi}, \eqref{eq:Xi1}, and \eqref{eq:Xi2}. 
They must also fulfill \eqref{conssin} and \eqref{consoct}, which follow from the conservation of the number of heavy quarks. 
Those equations, for instance, implement the fact that the decay width of the singlet is related with $\Xi_{os}$ because any singlet that decays contributes to the creation of an octet. 
As discussed in Appendix~\ref{E>m_D}, the final expressions for the $\Xi$'s can be cast, after a suitable redefinition of the density matrices,
in the following form:
\bea
\Xi_{so}(\rho_o,t) &=& \frac{2\,\als(\mu_E)T}{3N_c}(r_ih_o-h_sr_i)\rho_o(h_or_i-r_ih_s)\,,
\label{weak3}\\
\Xi_{os}(\rho_s,t) &=& \frac{4C_F\,\als(\mu_E)\,T}{3}(h_or_i-r_ih_s)\rho_s(r_ih_o-h_sr_i)\,,
\label{weak4}\\
\Xi_{oo}(\rho_o,t) &=& \frac{(N_c^2-4)\als(\mu_E)\,T}{3N_c}[h_o,r_i]\rho_o[r_i,h_o]\,,
\label{weak5}
\eea
which are suitable to be written in a Lindblad form. 

The Hermitian operator $H$ entering the Lindblad equation is 
\begin{equation}
H=\left(\begin{array}{c c}
h_{s} + {\rm Re}\,(-i\Sigma_s) & 0 \\
0 & h_{o} + {\rm Re}\,(-i\Sigma_o) 
\end{array}\right)\,.
\label{eq:H_Lindblad}
\end{equation}
Furthermore, the Lindblad equation is made in this case by a set of 9 collapse operators $C^0_i$, $C^1_i$, and $C^2_i$, which are
\begin{equation}
C^0_i=\sqrt{\frac{2\,\als(\mu_E)\,T}{3N_c}} \left[\frac{2ip_i}{M}+\frac{N_c\,\als(1/a_o)\,r_i}{2r}\right]
\left(\begin{array}{c c}
0 & 1 \\
0 & 0 
\end{array}\right)\,,
\end{equation}
\begin{equation}
C^1_i=\sqrt{\frac{4C_F\,\als(\mu_E)\,T}{3}} \left[-\frac{2ip_i}{M}+\frac{N_c\,\als(1/a_o)\,r_i}{2r}\right]
\left(\begin{array}{c c}
0 & 0 \\
1 & 0 
\end{array}\right)\,,
\end{equation}
\begin{equation}
C^2_i=\frac{2}{M}\sqrt{\frac{(N_c^2-4)\als(\mu_E)\,T}{3N_c}}p_i
\left(\begin{array}{c c}
0 & 0 \\
0 & 1
\end{array}\right)
\,,
\end{equation}
where $p_i=-i\partial/\partial x^i$.

To solve the Lindblad equation numerically we proceed along the lines discussed in Sec.~\ref{sec:spherical} 
and perform an expansion in spherical harmonics keeping only the $S$- and $P$-wave terms. The Hamiltonian reduces to
\begin{equation}
H=\left(\begin{array}{c c c c}
h_{s}^0 + {\rm Re}\,(-i\Sigma_s) & 0 & 0 & 0\\
0 & h_{s}^1 + {\rm Re}\,(-i\Sigma_s) & 0 & 0 \\
0 & 0 & h_{o}^0 + {\rm Re}\,(-i\Sigma_o) & 0 \\
0 & 0 & 0 & h_{o}^1 + {\rm Re}\,(-i\Sigma_o)
\end{array}\right)\,.
\end{equation}
The nine collapse operators above combine into three upon projection,
\begin{equation}
C^0=\sqrt{\frac{2\,\als(\mu_E)\,T}{3N_c}} 
\left(\begin{array}{c c c c}
0 & 0 & 0 & \frac{\frac{2}{M}\left(\frac{\partial}{\partial r} +\frac{2}{r}\right)+\frac{N_c\,\als(1/a_o)}{2}}{\sqrt{3}} \\
0 & 0 & \frac{2}{M}\frac{\partial}{\partial r} +\frac{N_c\,\als(1/a_o)}{2} & 0 \\
0 & 0 & 0 & 0 \\
0 & 0 & 0 & 0
\end{array}\right)\,,
\end{equation}
\begin{equation}
C^1=\sqrt{\frac{4C_F\,\als(\mu_E)\,T}{3}} 
\left(\begin{array}{c c c c}
0 & 0 & 0 & 0 \\
0 & 0 & 0 & 0 \\
0 & \frac{-\frac{2}{M}\left(\frac{\partial}{\partial r} +\frac{2}{r}\right)+\frac{N_c\,\als(1/a_o)}{2}}{\sqrt{3}} & 0 & 0 \\
-\frac{2}{M}\frac{\partial}{\partial r} +\frac{N_c\,\als(1/a_o)}{2} & 0 & 0 & 0
\end{array}\right)\,,
\end{equation}
\begin{equation}
C^2=-i\frac{2}{M}\sqrt{\frac{(N_c^2-4)\als(\mu_E)\,T}{3 N_c}}
\left(\begin{array}{c c c c}
0 & 0 & 0 & 0 \\
0 & 0 & 0 & 0 \\
0 & 0 & 0 & \frac{\frac{\partial}{\partial r} +\frac{2}{r}}{\sqrt{3}} \\
0 & 0 & \frac{\partial}{\partial r} & 0
\end{array}\right)\,.
\end{equation}
The operators $\partial/\partial r$ and $\partial/\partial r + 2/r$ come from  $p_i$ acting on $S$-wave and $P$-wave states, respectively.

\subsection{Results}
\label{sec:weak:results}
\begin{figure}[ht]
\begin{center}
\includegraphics[scale=0.4]{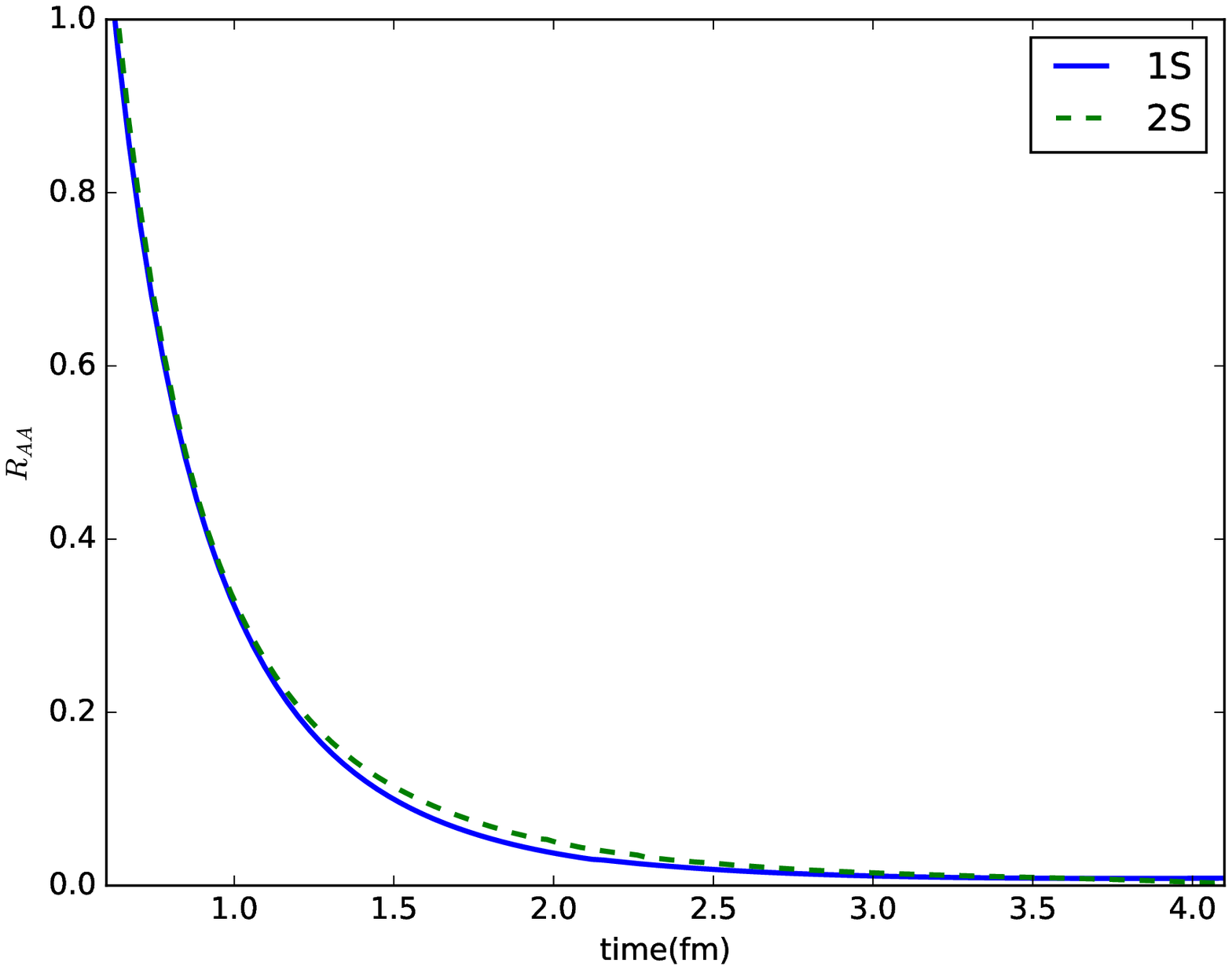}
\includegraphics[scale=0.4]{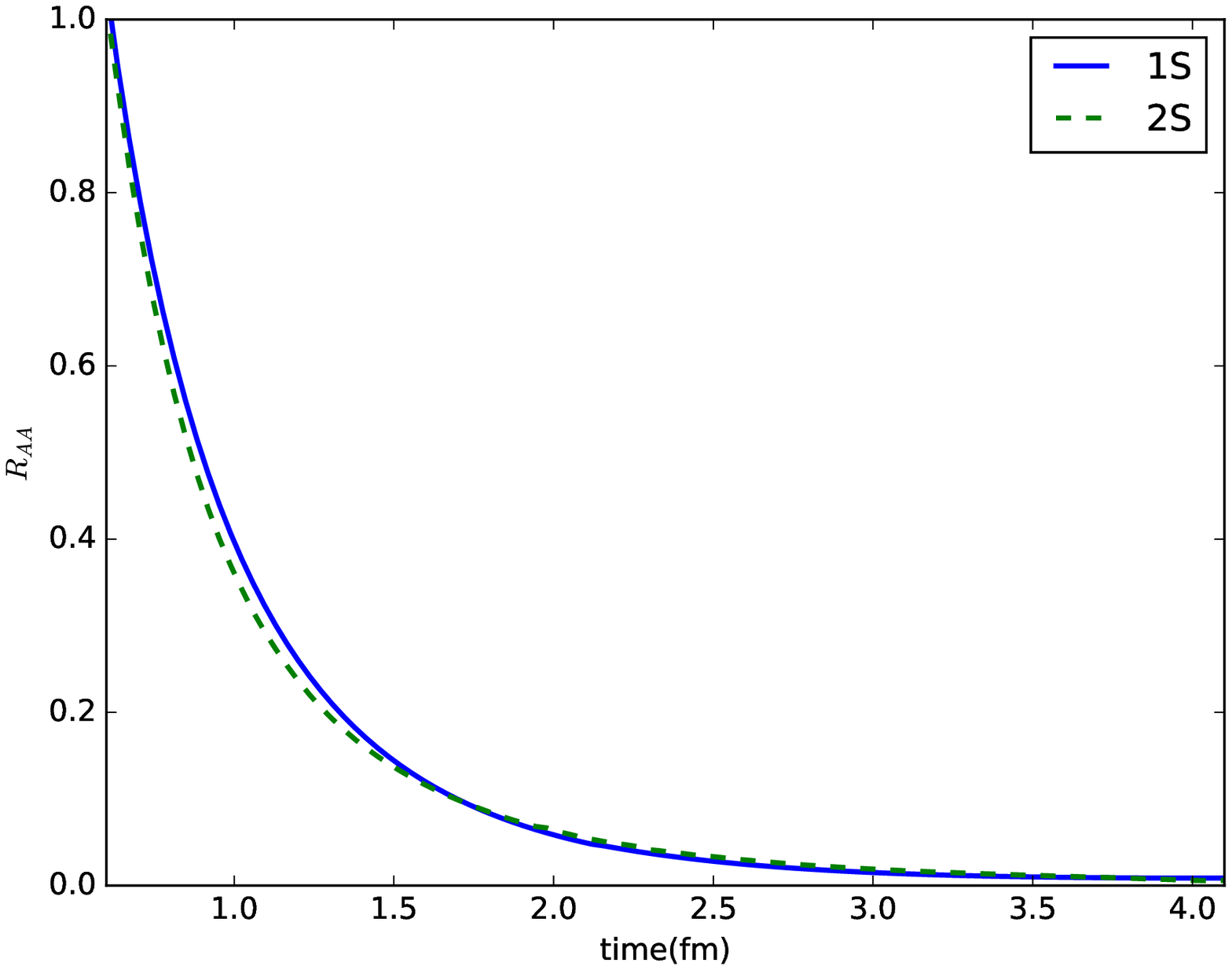}
\caption{Time evolution of $R_{AA}$ for bottomonium in the regime $1/a_0 \gg T \gg E \gg m_D$ with $\delta = 1$ and $\mu_E = 900$~MeV (left plot),  
  and $\mu_E = \pi T$  (right plot). We consider only the most central collisions ($b=0$ in \eqref{T0b}).
\label{fig:E}
}
\end{center}
\end{figure}

\begin{figure}[H]
\begin{center}
\includegraphics[scale=0.4]{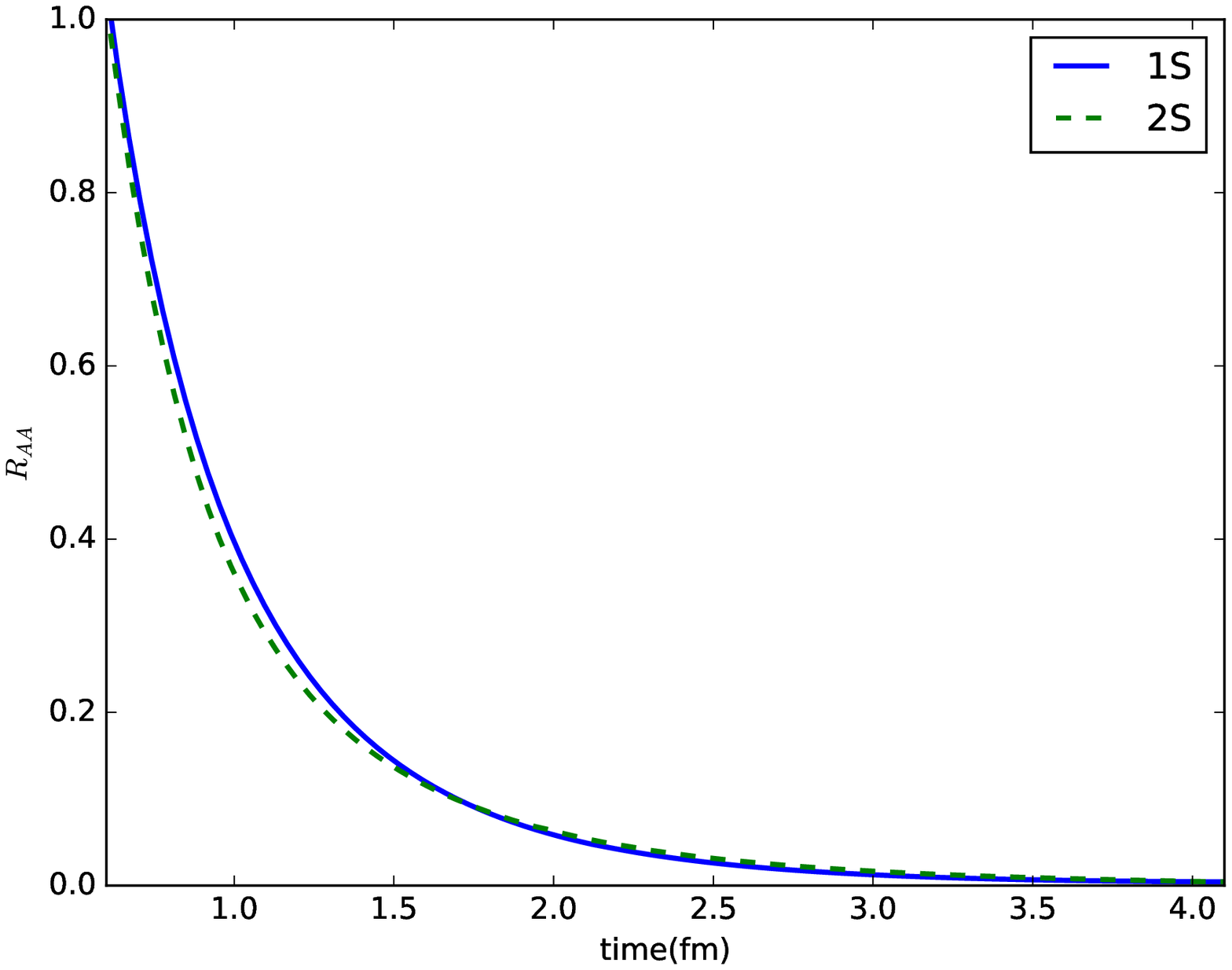}
\includegraphics[scale=0.4]{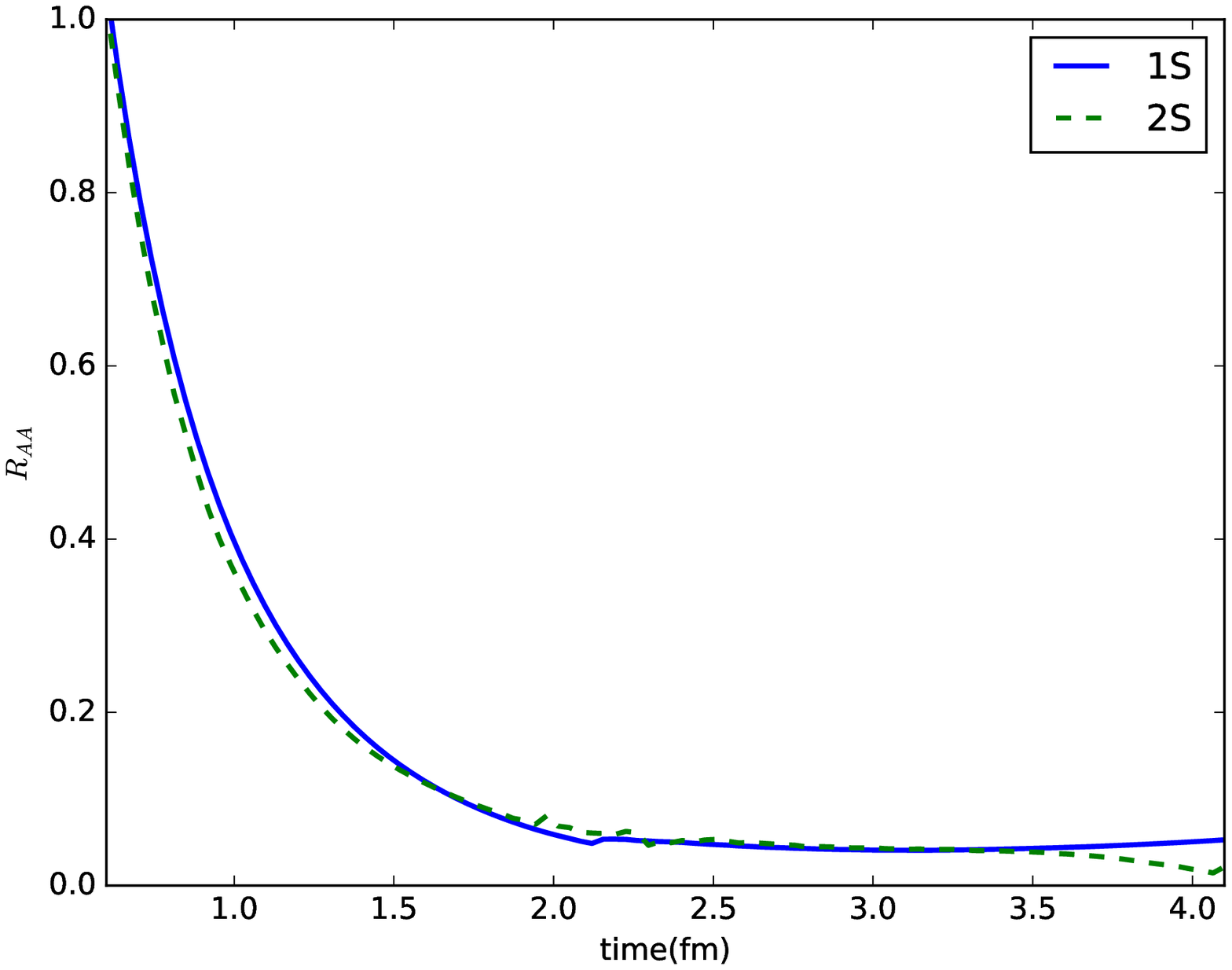}
\caption{Time evolution of $R_{AA}$ for bottomonium in the regime $1/a_0 \gg T \gg E \gg m_D$ with $\delta = 0.1$ (left plot) and $\delta = 10$ (right plot), and $\mu_E = \pi T$.
\label{E_running}
}
\end{center}
\end{figure}

In Fig.~\ref{fig:E} (left plot) we show the results that we obtain using $\mu_E = 900$~MeV (chosen to be the closest scale to~$-E_0$ where perturbation theory may still apply)
and the initial conditions as defined in Sec.~\ref{sec:initial} with $\delta = 1$. 
We see that the suppression is slightly larger for $\Upsilon(1S)$ than for $\Upsilon(2S)$ and strong for both.
This is in fact consistent with the leading order (LO) behavior of the decay widths calculated in \cite{Brambilla:2010vq}, but clashes with experimental observations. 
This can be due to the fact that one (or several) of the assumed hierarchies of energy scales may not be fulfilled
(e.g., if $T\sim m_D$, the plasma is not weakly coupled; see the next section) or that $\als$ is not small enough. 
In particular, for our choice of $\mu_E$, $\als(\mu_E)$ is rather large. 
Hence, it could happen that perturbation theory is still reliable at the scale $T$, but not at the scale $E$. 
If we repeat the computation using $\mu_E = \pi T$, we obtain the result shown in Fig.~\ref{fig:E} (right plot).
We see now that the suppressions of $\Upsilon(1S)$ and $\Upsilon(2S)$ are similar, but still very strong, both features are difficult to reconcile with experimental observations.
Finally, we investigate the sensitivity to the initial conditions.
We show in the left and right plots of  Fig.~\ref{E_running} the results for $\delta = 0.1$ (more singlets than in the LO NRQCD production)
and $\delta = 10$ (more octets than in the LO NRQCD production), respectively. 
While the case $\delta = 0.1$ is similar to the case $\delta = 1$, for $\delta = 10$ we observe slightly less quarkonium suppression.
This indicates that above some threshold the suppression pattern will be quite sensitive to the ratio between the singlets and the octets initially produced.
In particular, the larger the initial fraction of the quark-antiquark color octets is, the more marked is the feedback of singlets coming from the octet decays.
This is at the origin of the small kink at about 2~fm in the right plot of Fig.~\ref{E_running}.

\section{Quarkonium in a strongly-coupled plasma: $1/a_0 \gg T \sim m_D \gg E$}
\label{sec:strong}
In this section we apply the general equations derived in Sec.~\ref{sec:tevol} to the case in which the thermodynamical scales are smaller than $1/a_0$ but larger than the binding energy $E$.  
Thermodynamical scales are $T$ and the Debye mass $m_D$.
In a weakly-coupled plasma one assumes that $T \gg m_D \sim gT$. 
In this section, however, we assume more generally that the plasma is strongly coupled. This means that we take $T \sim m_D$.
As we will see, the evolution equations can be written in terms of just two real constants~\cite{Brambilla:2016wgg}. 
Because the entire information about the medium is contained in these two constants that we are not going to evaluate, 
we do not need to make here any special assumption about the properties and degrees of freedom of the medium 
(which in the previous section was explicitly taken to be a quark-gluon plasma).
Hence, everything that we write in this section applies to a generic strongly-coupled hot medium. 

If $1/a_0 \gg T$, $\Lambda_{\rm QCD}$, we can describe the evolution of the heavy-quark densities in the fireball by means of the equations found in Sec.~\ref{sec:tevol}.  
If $T$ is larger than $E$, we may neglect the energy-dependent exponentials $e^{\pm ih_{s,o}(t-t_2)}$ that appear in the definitions of the $\Sigma$'s and $\Xi$'s:
for $t-t_2\sim 1/T$, $e^{\pm ih_{s,o}(t-t_2)}\sim 1$ holds.
Using also the approximations \eqref{approx1} and \eqref{approx2} (this is where the quasistatic approximation enters), we can write 
\begin{eqnarray}
\Sigma_s(t) &=& \frac{r^2}{2}\left[ \kappa(t) + i \gamma(t) \right]\,,
\label{strong1}
\\
\Sigma_o(t) &=& \frac{N_c^2-2}{2(N_c^2 -1)}\frac{r^2}{2}\left[ \kappa(t) + i \gamma(t) \right]\,,
\label{strong2}
\\
\Xi_{so}(\rho_o,t) &=& \frac{1}{N_c^2 -1}\,r^i \, \rho_o \, r^i\,\kappa(t)\,,
\label{strong3}
\\
\Xi_{os}(\rho_s,t) &=& r^i \, \rho_s \, r^i\,\kappa(t)\,,
\label{strong4}
\\
\Xi_{oo}(\rho_o,t) &=& \frac{N_c^2-4}{2(N_c^2 -1)}\,r^i \, \rho_o \, r^i\,\kappa(t)\,.
\label{strong5}
\end{eqnarray}

The real quantity $\kappa$ is the heavy-quark momentum diffusion coefficient~\cite{CasalderreySolana:2006rq,CaronHuot:2007gq}:
\begin{equation}
\kappa  \equiv  \frac{g^2}{6\,N_c} \,  {\rm Re} \int_{-\infty}^{+\infty}ds \, \langle T\,E^{a,i}(s,{\bf 0}) E^{a,i}(0,{\bf 0})\rangle
= \frac{g^2}{6\,N_c} \, \int_0^{\infty} ds \, \langle \{E^{a,i}(s,{\bf 0}), E^{a,i}(0,{\bf 0})\}\rangle\,,
\label{kappa}
\end{equation} 
where $T$ stands for time ordering. 
The quantity $\kappa$ is related to the thermal decay width of the heavy quarkonium. 
In particular for $1S$ states, we have  (see \eqref{gammaself})
\begin{equation}
\Gamma(1S) =  -2 \langle {\rm Im}\,(-i\Sigma_s) \rangle = 3 a_0^2 \,\kappa \,.
\label{width}
\end{equation}
The heavy-quark momentum diffusion coefficient has been recently computed on the lattice~\cite{Francis:2015daa}: 
\begin{equation}
1.8 \lesssim \frac{\kappa}{T^3} \lesssim 3.4 \,.
\label{kapparange}
\end{equation}
The above estimate has been obtained from pure SU(3) calculations at temperatures of about 1.5~$T_c$.
Higher temperatures seem to suggest a smaller $\kappa$ and/or a temperature dependence of $\kappa/T^3$~\cite{Banerjee:2011ra}.
Also light quarks may modify the above values.
A perturbative determination of $\kappa$ at next-to-leading order can be found in~\cite{CaronHuot:2008uh}.
At next-to-leading order $\kappa/T^3$ turns out to be in the range $9 \lesssim \kappa/T^3 \lesssim 18$, 
but, as pointed out in~\cite{CaronHuot:2008uh}, the perturbative series does not converge for any realistic value of the strong coupling.
In fact, at leading order (see \eqref{eq:vsp}) it turns out to deliver a negative $\kappa$ for realistic LHC temperatures. 
At the moment no definite statement can be drawn from looking at $\kappa$ in perturbation theory.

The real quantity $\gamma$ is defined as 
\begin{equation}
\gamma  \equiv  \frac{g^2}{6\,N_c} \, {\rm Im}\int_{-\infty}^{+\infty}ds \, \langle T\,E^{a,i}(s,{\bf 0}) E^{a,i}(0,{\bf 0})\rangle
= -i \frac{g^2}{6\,N_c} \, \int_0^{\infty} ds \, \langle [E^{a,i}(s,{\bf 0}), E^{a,i}(0,{\bf 0})]\rangle\,.
\label{gamma}
\end{equation}
The quantity $\gamma$ is related to the thermal mass shift of the heavy quarkonium. 
In particular for $1S$ states, we have (see \eqref{deltam}) 
 \begin{equation}
\delta M(1S) =  \langle {\rm Re}\,(-i\Sigma_s) \rangle = \frac{3}{2} a_0^2 \,\gamma \,, 
\label{mass}
\end{equation}
So far $\gamma$ has not been computed on the lattice.
The only estimate we have for $\gamma$ is the perturbative calculation done at leading order in~\cite{Brambilla:2008cx}; see \eqref{gammapert}.
Note that in our setup both $\kappa$ and $\gamma$ depend on time through the evolving temperature of the plasma.

With the above functions, we can write the evolution equations \eqref{eq:ev} and \eqref{eq:ev_octet} in the Lindblad form \eqref{eq:Lindblad}.
In addition to the Hermitian Lindblad operator $H$, 
\begin{equation}
H = \left(\begin{array}{c c}
h_s & 0\\
0 & h_o
\end{array}\right)
+ \frac{r^2}{2}\,\gamma(t)\, 
\left(\begin{array}{c c}
1 & 0\\
0 & \frac{N_c^2-2}{2(N_c^2-1)}
\end{array}\right)\,,
\label{Hlindstrong}
\end{equation}
we need six collapse operators $C^0_i$ and $C^1_i$, which are
\begin{equation}
C^0_i=\sqrt{\frac{\kappa(t)}{N_c^2-1}}\,r^i\left(\begin{array}{c c}
0 & 1\\
\sqrt{N_c^2-1} & 0
\end{array}\right)\,,
\end{equation}
\begin{equation}
C^1_i=\sqrt{\frac{(N_c^2-4)\kappa(t)}{2(N_c^2-1)}}\,r^i\left(\begin{array}{c c}
0 & 0\\
0 & 1
\end{array}\right)\,.
\label{lindstrong}
\end{equation}
Following Sec.~\ref{sec:spherical}, the numerical computation is done by expanding the density matrix and the collapse operators in spherical harmonics, and keeping only $S$- and $P$-waves.
The projected Hamiltonian reads
\begin{equation}
H = \left(\begin{array}{c c c c}
h_s^0 & 0 & 0 & 0\\
0 & h_s^1 & 0 & 0\\
0 & 0 & h_o^0 & 0 \\
0 & 0 & 0 & h_o^1
\end{array}\right)
+ \frac{r^2}{2}\,\gamma(t)\, 
\left(\begin{array}{c c c c}
1 & 0  & 0 & 0\\
0 & 1 & 0 & 0\\
0 & 0 & \frac{N_c^2-2}{2(N_c^2-1)} & 0 \\
0 & 0 & 0 &\frac{N_c^2-2}{2(N_c^2-1)}
\end{array}\right)\,,
\end{equation}
and the six collapse operators above are combined into two,
\begin{equation}
C^0 = \sqrt{\frac{\kappa(t)}{N_c^2-1}}\,r\left(\begin{array}{c c c c}
0 & 0 & 0 & \frac{1}{\sqrt{3}}\\
0 & 0 & 1 & 0 \\
0 & \frac{\sqrt{N_c^2-1}}{\sqrt{3}} & 0 & 0 \\
\sqrt{N_c^2-1} & 0 & 0 & 0
\end{array}\right)\,,
\end{equation}
\begin{equation}
C^1 = \sqrt{\frac{(N_c^2-4)\kappa(t)}{2(N_c^2-1)}}\,r\left(\begin{array}{c c c c}
0 & 0 & 0 & 0\\
0 & 0 & 0 & 0\\
0 & 0 & 0 & \frac{1}{\sqrt{3}}\\
0 & 0 & 1 & 0
\end{array}\right)\,.
\end{equation}

\subsection{Results}
\label{sec:strong:results}
\begin{figure}[ht]
  \includegraphics[scale=0.4]{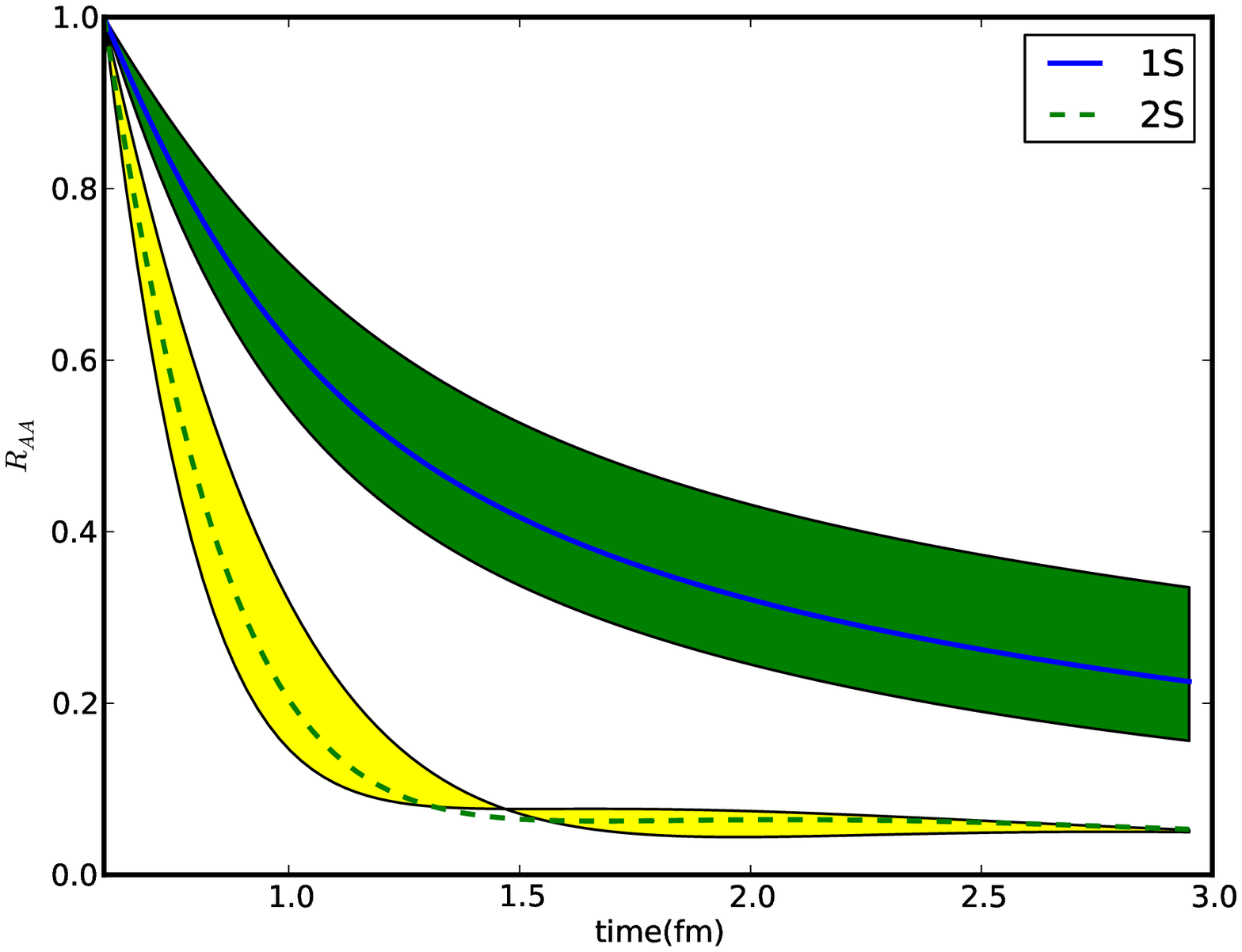}
  \includegraphics[scale=0.4]{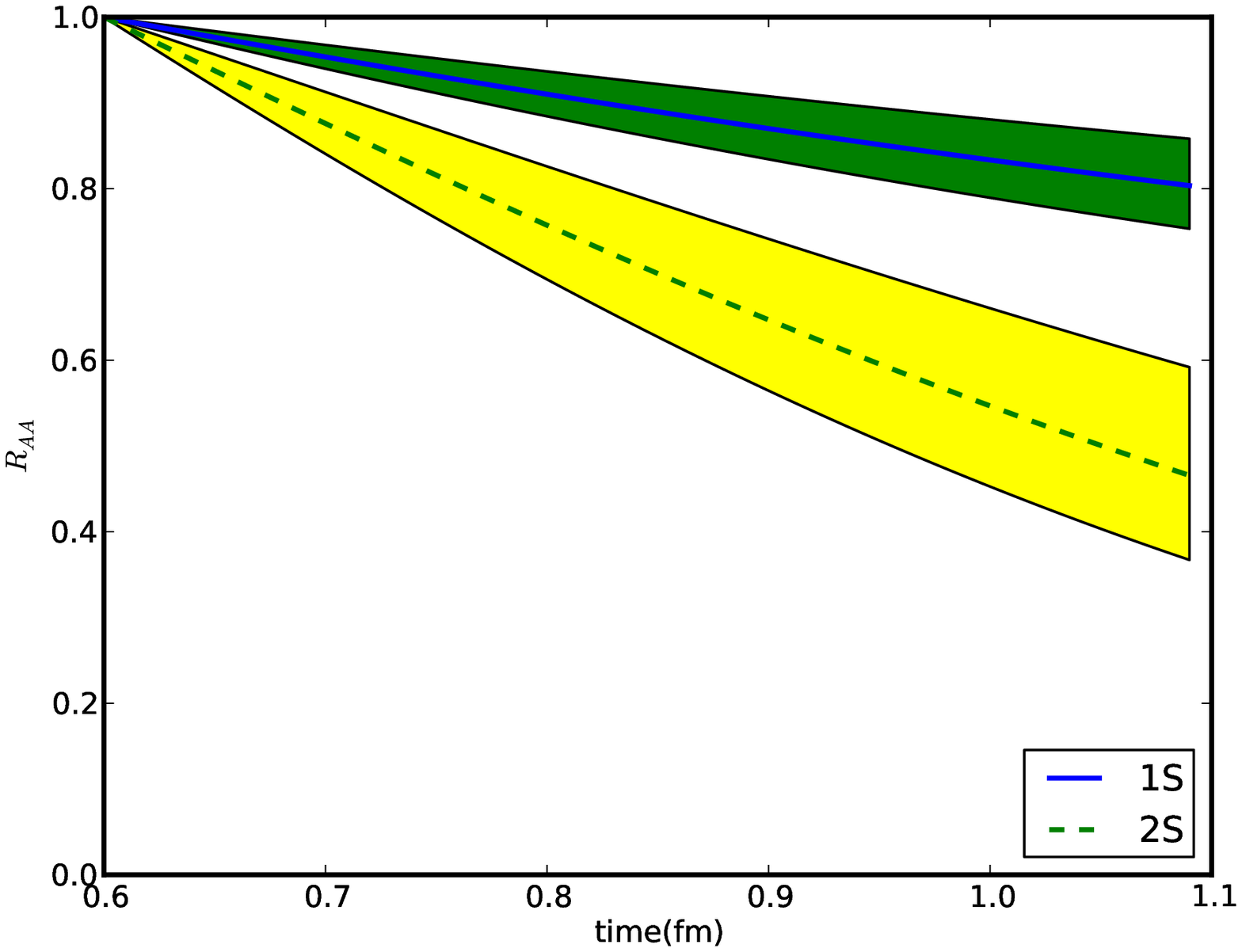}
\caption{Time evolution of $R_{AA}$ for bottomonium with $\kappa/T^3$ in the range \eqref{kapparange}, $\gamma = 0$ and $\delta =1$, 
  for 30-50\% centrality (left plot) and for 50-100\% centrality (right plot).
\label{fig:ev}
}
\end{figure}

In the bottomonium case, the time evolutions of $R_{AA}$ for 30-50\% centrality and 50-100\% centrality are shown in the left and right plots of Fig.~\ref{fig:ev}, respectively. 
Note that, in the left plot, the $R_{AA}$ for the $2S$ state becomes insensitive to $\kappa$ at large times,
an indication that it reaches a steady state before the quark-gluon plasma vanishes.

We have taken $\kappa/T^3$ in the range \eqref{kapparange}, while we have set $\gamma = 0$ and $\delta =1$.
We have no \textit{a priori} knowledge for either $\gamma$, since a non-perturbative determination of this parameter is missing, 
or $\delta$, since we ignore the precise initial conditions.
Hence, we have scanned for several values of $\gamma$ and $\delta$. Here, we have restricted ourselves to $\gamma <0$, the sign of the perturbative result~\eqref{gammapert}.
We find that the CMS results of~\cite{Chatrchyan:2012lxa} prefer small values of $\gamma$, 
which is the rationale for choosing $\gamma = 0$. 
We further find that the results are rather insensitive to $\delta$, in contrast to the weakly-coupled case discussed in the previous section. 
The choice $\delta =1$ assumes the initial ratio of octets over singlets to be just $1/\als(M)$.

\begin{table}[ht]
\begin{center}
\begin{tabular}{|c|c|c|c|c|c|}\hline 
\multicolumn{2}{|c|}{30-40\% centrality}&\multicolumn{2}{|c|}{40-50\% centrality}&\multicolumn{2}{|c|}{50-70\% centrality} \\
\hline
$R_{AA}(1S)$ & $\frac{R_{AA}(2S)}{R_{AA}(1S)}$ & $R_{AA}(1S)$ & $\frac{R_{AA}(2S)}{R_{AA}(1S)}$ & $R_{AA}(1S)$ & $\frac{R_{AA}(2S)}{R_{AA} (1S)}$ \\
\hline
$0.20^{+0.10}_{-0.06}$ & $0.25^{+0.11}_{-0.09}$ & $0.27^{+0.11}_{-0.13}$ & $0.21\pm 0.08$ & $0.47^{+0.10}_{-0.08}$ & $0.10^{+0.04}_{-0.01}$ \\
\hline
\end{tabular}
\end{center}
\caption{Results for $R_{AA}(1S)$ and $R_{AA}(2S)$ for $\kappa/T^3$ in the range \eqref{kapparange}, $\gamma = 0$ and $\delta =1$ in the bottomonium case.
\label{tab:results}}
\end{table}

\begin{figure}[ht]
\includegraphics[scale=0.4]{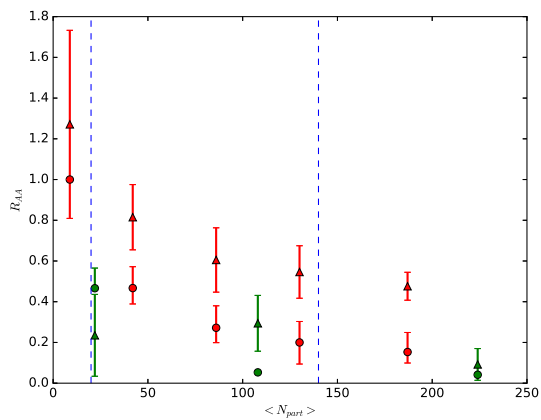}
\caption{$R_{AA}$ as obtained from Table~\ref{tab:results} (dots) compared with the CMS data of~\cite{Khachatryan:2016xxp} (triangles).
Upper (red) entries refer to the $\Upsilon(1S)$, and lower (green) entries to the $\Upsilon(2S)$. 
The vertical dashed lines highlight
 the window in which we expect the approximation $1/a_0 \gg T\sim m_D\gg E$ to be valid.
\label{fig:RAA}
}
\end{figure}

In Table~\ref{tab:results} we show our predictions for the centrality bins studied by CMS at 2.76~TeV in ~\cite{Khachatryan:2016xxp}.
Results in this table are corrected for feed-down effects using the method of~\cite{Strickland:2011aa} with the updated feed-down fractions from \cite{Krouppa:2015yoa}.
The reason why feed-down is taken into account in the table and not in the time evolution plots is that it takes place after freeze-out. 
In order to be on the safe side regarding the condition $1/a_0 \gg T \sim m_D \gg E$, we focus only on centralities between $30 \%$ and $70 \%$. 
All our determinations are summarized and compared with the CMS data in Fig.~\ref{fig:RAA}.
The theoretical error band accounts only for the lattice uncertainty in $\kappa$.

$\Upsilon$ suppression in heavy-ion collisions has also been studied by the Alice Collaboration~\cite{Abelev:2014nua}.
They have only considered the centrality bins $0-20 \%$ and $20-90 \%$.
The initial temperature for the $0-20 \%$ centrality bin is too high for our present study.
Regarding the centrality bin $20-90 \%$, the average initial temperature happens to be very similar to the one in the centrality bin $50-70 \%$
and, therefore, our prediction is approximately the same.
Analyses for LHC data at 5.02~TeV are under way~\cite{Fronze:2016gsr}.

\begin{figure}[ht]
  \includegraphics[scale=0.5]{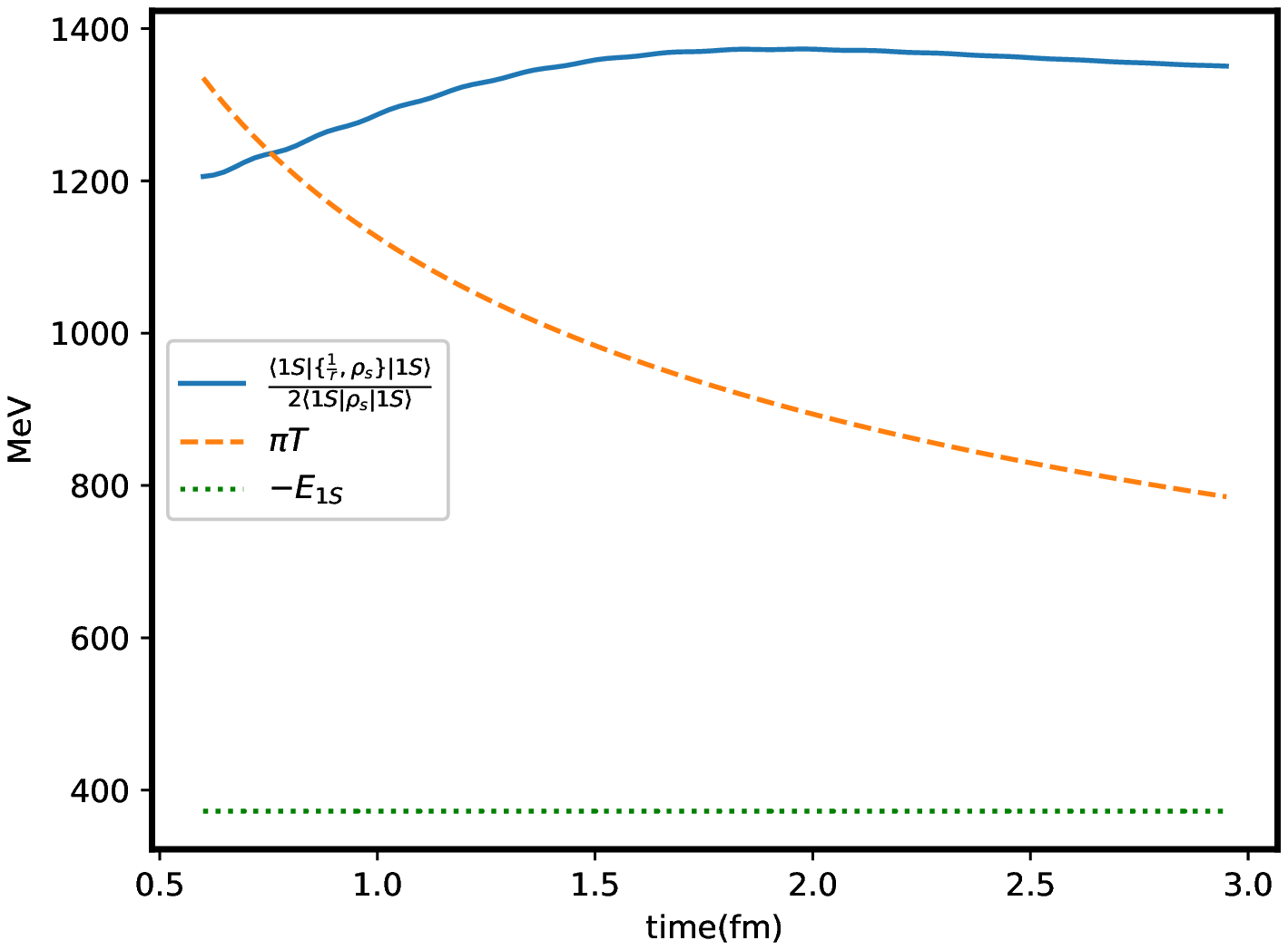}
  \includegraphics[scale=0.5]{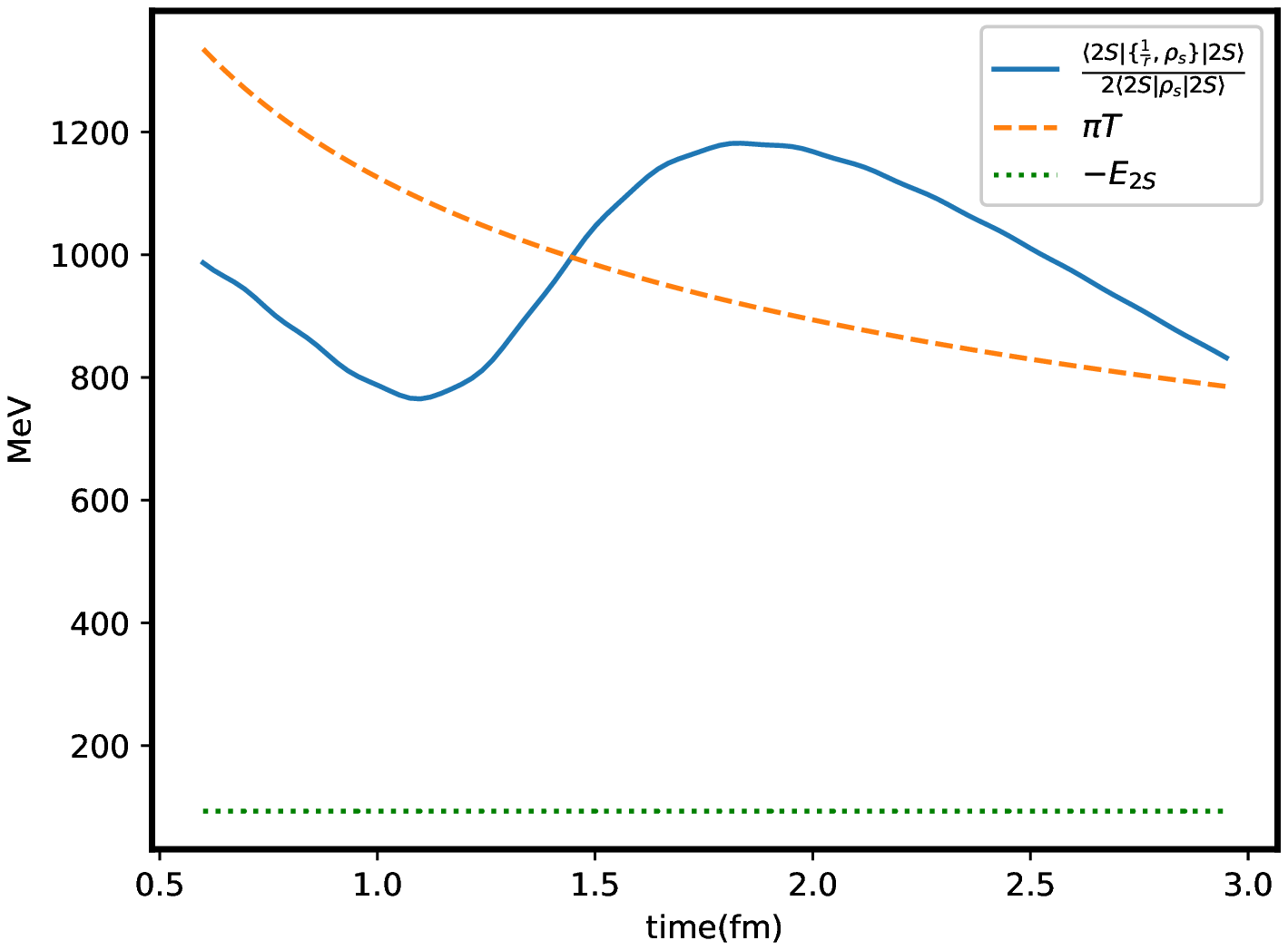}
\caption{Time evolution of $\langle nS| \{1/r,\rho_s\}/2| nS\rangle$ (continuous line), the binding energy $E_{nS}$ (dotted line), and the temperature (dashed line) for bottomonium $n=1$ (left plot) and $n=2$ states (right plot), with $\kappa/T^3 = 2.6$, $\gamma = 0$, $\delta =1$, and for 30-50\% centrality collisions.
\label{fig:rE}
}
\end{figure}

In order to check that the hierarchy of scales assumed at the beginning of Sec.~\ref{sec:strong} is maintained during the evolution of the $\Upsilon$ states in the fireball, we have computed the time evolution of the matrix element $\langle nS| \{1/r,\rho_s\}/2| nS \rangle$ and of the binding energy $E_{nS}$ for bottomonium $n=1$ and $n=2$ states. This is shown in Fig.~\ref{fig:rE}.
We see that both $\langle 1S| \{1/r,\rho_s\}/2| 1S\rangle/\langle 1S| \rho_s| 1S\rangle$
and $\langle 2S| \{1/r,\rho_s\}/2| 2S\rangle/\langle 2S| \rho_s| 2S\rangle$
remain close to their initial, in-vacuum, values: $1.334$~GeV and $0.884$~GeV respectively.
For our choice of parameter $\gamma=0$, the Hamiltonian $H$ coincides with the in-vacuum one, see \eqref{Hlindstrong}, and so do the binding energies:
$E_{1S} \approx 0.37$~GeV and $E_{2S} \approx 0.04$~GeV.
This implies that the in-vacuum hierarchy between the inverse sizes of the bound states and their binding energies is preserved during the evolution in the plasma.
Furthermore, we also plot $\pi T$ as a function of time for the $30-50\%$ centrality class.
We observe that the hierarchy between the bound-state scales and the thermal scale is preserved for late time evolution.

\begin{figure}[ht]
\includegraphics[scale=0.4]{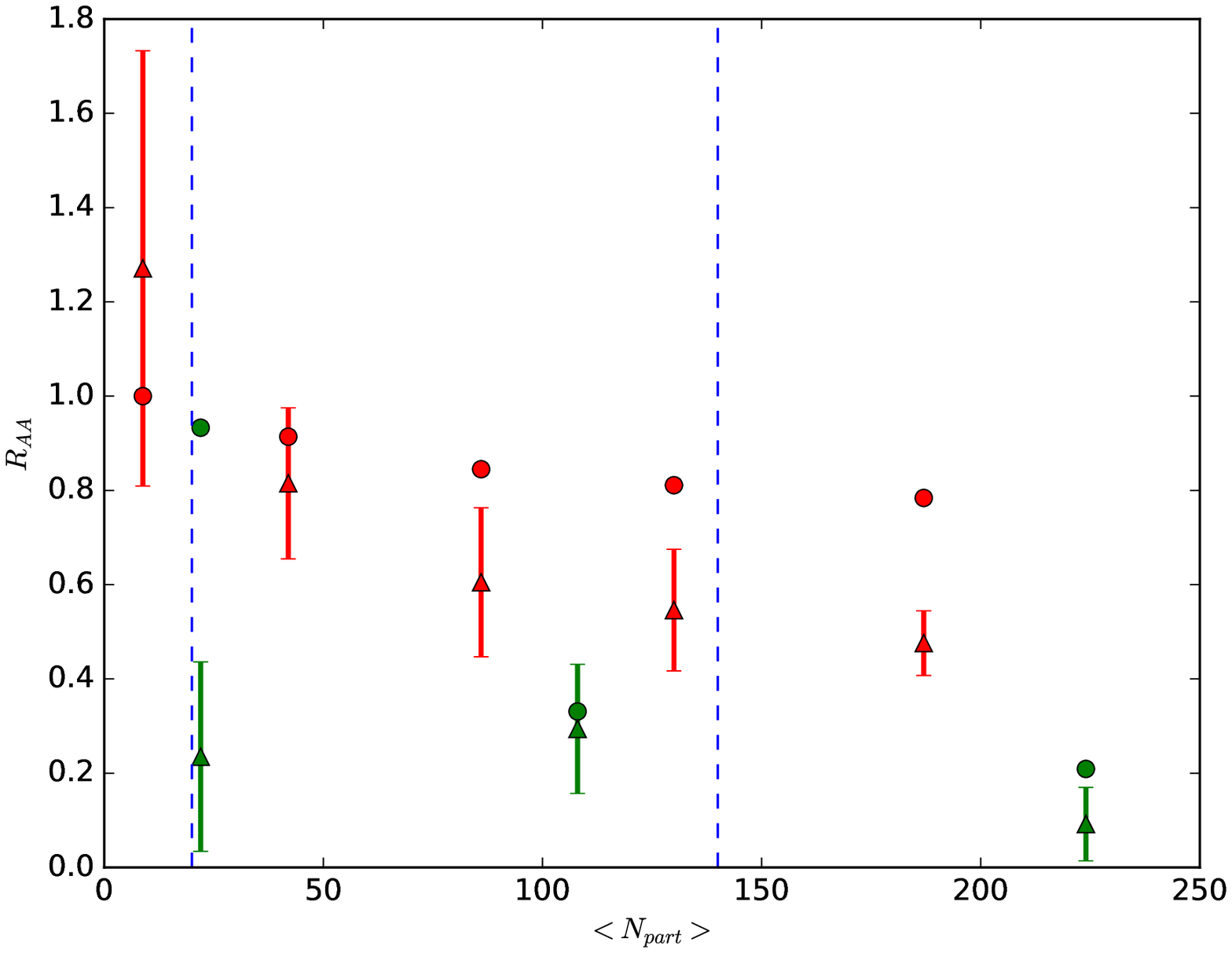}
\includegraphics[scale=0.4]{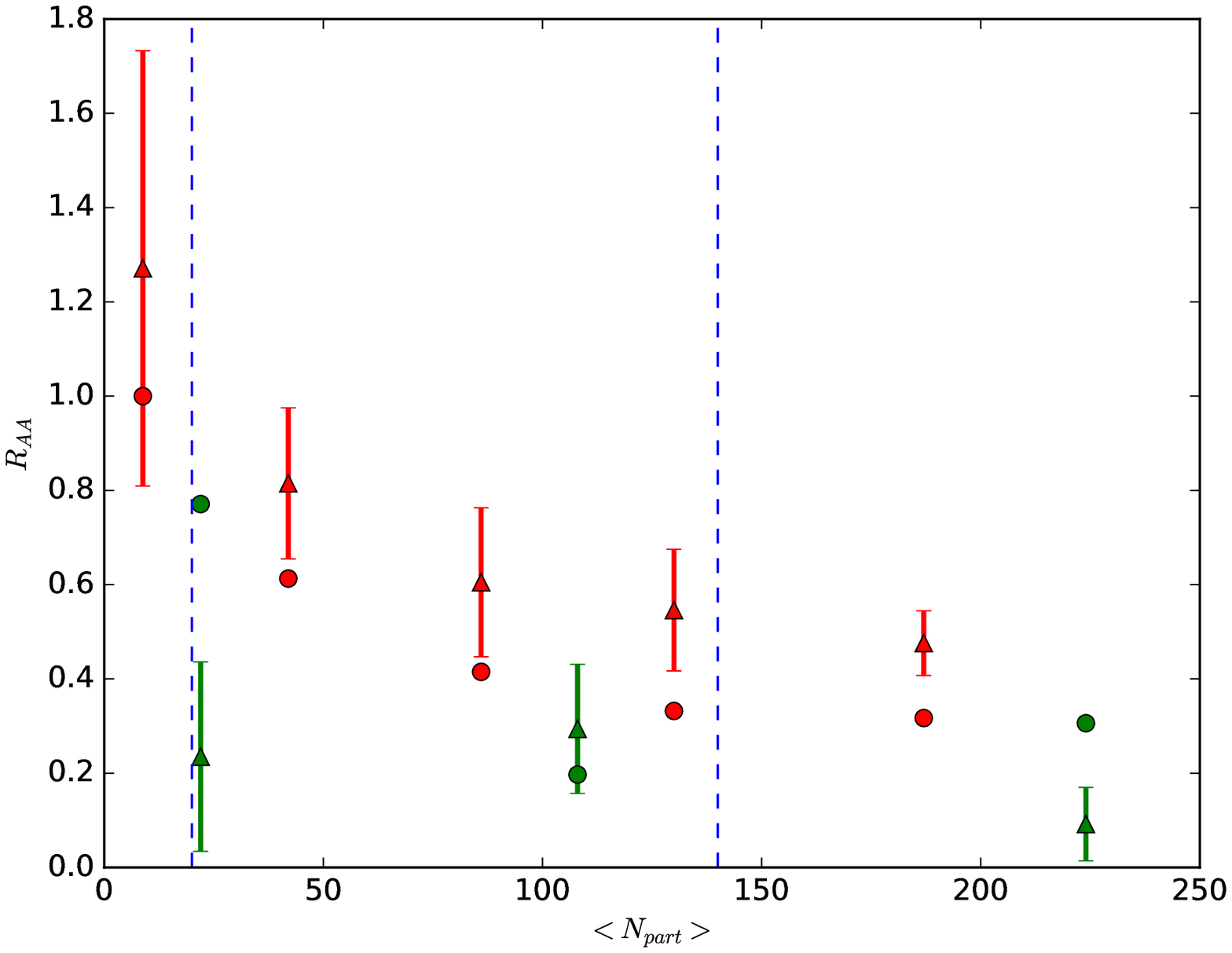}
\caption{$R_{AA}$ as obtained using $\kappa/T^3 = 0.25$ and $\gamma=0$ in the left plot and $\kappa/T^3 = 2.6$ and $\gamma/T^3 = 6$ in the right plot (dots)
compared with the CMS data of~\cite{Khachatryan:2016xxp} (triangles).
Upper (red) entries refer to the $\Upsilon(1S)$, and lower (green) entries to the $\Upsilon(2S)$. 
The vertical dashed lines highlight the window in which we expect the approximation $1/a_0 \gg T\sim m_D\gg E$ to be valid.
\label{fig:RAA_k025gamma6}
}
\end{figure}

Many effects that have not been considered in the present analysis or considered in a simplified form (e.g., the hydrodynamical evolution) 
have the potential to quantitatively impact the nuclear modification factor calculated here. 
Inside the framework presented here, the results depend on the initial conditions and on just the parameters $\kappa$ and $\gamma$. 
The impact of a value of $\kappa$ outside the range in \eqref{kapparange} or of a positive value of $\gamma$ is shown for two illustrative examples in Fig.~\ref{fig:RAA_k025gamma6}.
The result suggests that there may exist values of $\kappa$, $\gamma$, and $\delta$ that reproduce the present data.
The fact that a value of $\kappa$ lower than the lattice prediction is needed may be explained
if most of the quark-antiquark pairs are moving with respect to the plasma~\cite{Escobedo:2011ie,Escobedo:2013tca}, at least in the weak-coupling case.
As mentioned before, the lattice results of \cite{Banerjee:2011ra} also seem to point to a lower value of kappa at higher temperatures.
Note finally that a positive value of $\gamma$ means that the medium is very different from a weakly-coupled quark-gluon plasma, since the latter has a negative gamma.
This reinforces the need for a lattice evaluation of $\gamma$.

From the comparison between strongly- and weakly-coupled results, it can be seen that there is more suppression in the weakly-coupled case than in the strongly-coupled one, 
which might seem surprising. In order to understand this, one has to take into account the several differences between the two cases. 
In the weakly-coupled case, we restrict ourselves to very central collisions (in order to guarantee high temperatures and make the weakly-coupled case plausible), 
while in the strongly-coupled scenario, which is more phenomenologically oriented, we focus on less central collisions since there our approximations are more likely to be fulfilled. 
Another important difference is that the leading-order thermal corrections in the weakly-coupled case are approximately linear with the temperature, 
while in the strongly-coupled one they are cubic. 
In a system that expands following Bjorken evolution and with a sound velocity close to the ideal gas case, linear corrections tend to have a larger impact than cubic ones. 
This can be seen by looking at our analysis of the static limit in Appendix~\ref{app1}.

\section{Conclusions and Outlook}
\label{sec:con}
In the paper we present a systematic description in an effective field theory framework (pNRQCD) of heavy quark-antiquark systems as open quantum systems 
interacting with an environment made of light quarks and gluons, the fireball formed in heavy-ion collisions.
While this work derives its original inspiration from~\cite{Akamatsu:2014qsa}, 
it is quite different from all previous studies on the subject~\cite{Young:2008he,Young:2010jq,Borghini:2011yq,Borghini:2011ms,Blaizot:2015hya,Akamatsu:2011se,Katz:2015qja,DeBoni:2017ocl,Kajimoto:2017rel}, 
as the derived evolution equations fulfill three essential conditions:
they conserve the total number of heavy quarks (i.e., $\Tr \{\rho_s\} + \Tr \{\rho_o\}$ is preserved by the evolution equations);
they account for the non-Abelian nature of QCD (through gluon exchanges color-singlet quarkonia may dissociate
into quark-antiquark color-octet states, and vice versa quark-antiquark color-octet states may generate quarkonia);
and, finally, they do not rely on classical approximations but rather follow from the closed-time-path formalism applied to quantum field theory.
The work substantially extends, updates, and completes a previous strongly-coupled analysis done in~\cite{Brambilla:2016wgg}.
A recent study also based on pNRQCD can be found in~\cite{Yao:2017fuc}.

The evolution equations in terms of quark-antiquark color-singlet and color-octet density matrices have been written in~\eqref{eq:ev} and~\eqref{eq:ev_octet}.
The equations rely on the assumption that the typical inverse size of the quark-antiquark system, $1/a_0$, is larger than any other scale of the medium and larger than $\Lambda_{\rm QCD}$. 
This implies that the quark-antiquark interaction is mainly Coulombic and that the interaction with the medium may be multipole expanded.
The evolution equations follow from the calculation of the singlet and octet density matrices at the leading non-trivial order in the multipole expansion.
Since the heavy quark density is expected to be small, we only keep linear terms in them.
Then we show that, at the order in the multipole expansion we are working,
the time derivative of the density matrices at a given time can be written as a linear function of the density matrices at the same time.
This produces an evolution equation that is reliable at arbitrary large times and that turns out to be equivalent to the Schwinger--Dyson equations depicted in Fig.~\ref{fig:resum}.
The evolution equations~\eqref{eq:ev} and~\eqref{eq:ev_octet} do not make any special assumption on the medium and may be valid 
either for a quark-gluon plasma or a different medium formed in the heavy-ion collisions.
They are also valid either if the medium is in thermal equilibrium or if it is far from it (provided that no dynamical scale is larger than $1/a_0$, as mentioned above).
The evolution equations preserve the number of heavy quarks, but, in general, they are not of the Lindblad form.

In order to get a Lindblad equation we consider some specific cases where we assume that the time scales we are interested in are larger than any other time scale in the problem
and that the evolution is quasistatic. For a quasistatic evolution the environment may be locally in thermal equilibrium.
Thermal equilibrium allows to define at each time a temperature.
In Sec.~\ref{sec:weak} we consider the explicit case of a weakly-coupled quark-gluon plasma. 
The temperature, $T$, and the Debye mass, $m_D$, are assumed to be such that $T \gg E \gg m_D$,
where $E$ is the typical binding energy of the heavy quark-antiquark system. 
In Sec.~\ref{sec:strong} we consider a strongly-coupled plasma at a temperature $T \sim m_D \gg E$.
In Appendix \ref{m_D>E} we also discuss the case $T\gg m_D \gg E$. 
In order to numerically solve the Lindblad equation, we further make a truncation on the number of partial waves taken into account.
This approximation is a of technical nature, and it is only useful to simplify the solution in the face of limited computing time.
In the future, one may relax some of these approximations, or even solve directly the evolution equations~\eqref{eq:ev} and~\eqref{eq:ev_octet}.

The numerical solutions of the Lindblad equations are presented and discussed in Sec.~\ref{sec:weak:results} for the weakly-coupled quark-gluon plasma, 
and in Sec.~\ref{sec:strong:results} for the strongly-coupled medium.
We analyze some specific set of initial conditions, parameter and time evolution of the thermal medium (we work with the Bjorken evolution \eqref{Tbj}).
A more extensive study is surely due in the future.
For instance, the Bjorken evolution could be substituted by more refined hydrodynamical models (see \cite{Yan:2017bgc} for a review of the state-of-the-art).
There is also plenty of room for improvement at early times, where color glass condensate physics~\cite{Gelis:2010nm,Gelis:2012ri}
may be included or higher-order NRQCD production results (see~\cite{Brambilla:2014jmp} and references therein) for the initial conditions of the density matrices may be incorporated. 
Eventually, the momentum dependence of the quark-antiquark pairs should also be addressed. This requires enlarging the Hilbert space of the density matrices and incorporating
the effects of the relative motion of the pair with respect to the medium; see, for instance~\cite{Escobedo:2011ie,Escobedo:2013tca}. 

The strongly-coupled plasma case may be of particular interest at the LHC, since the temperature $T$ or even $\pi T$ may not be much larger than $m_D$.
The Lindblad equation for the strongly-coupled case depends on only two parameters:
the heavy-quark momentum diffusion coefficient $\kappa$, defined in \eqref{kappa}, and $\gamma$, defined in \eqref{gamma}.
While $\kappa$ has been computed on lattice QCD (although in pure gluodynamics and for a limited range of temperatures), $\gamma$ has not.
A first determination of $\gamma$ remains therefore the most urgent missing ingredient for the computation of the suppression factor $R_{AA}$.

\acknowledgments
J.S. thanks Anton Andronic for providing invaluable information on the current codes used by the experimental collaborations to map the number of participants to average impact parameters.
The work of N.B. and A.V. was supported by the DFG Grant No. BR 4058/1-2 ``Effective Field theories for heavy probes of hot plasma''
and by the DFG cluster of excellence ``Origin and structure of the universe'' (www.universe-cluster.de). 
The work of M.A.E. was supported by the Academy of Finland, project 303756.
J.S. was supported by the CPAN  CSD2007-00042 Consolider--Ingenio 2010 program; 
and the FPA2013-43425-P, FPA2013-4657, FPA2016-81114-P and FPA2016-76005-C2-1-P projects (Spain); and the 2014-SGR-104 grant (Catalonia).

\appendix

\section{Static limit}
\label{app1}
In this appendix, we consider the evolution equations for static quarks under the condition 
\begin{equation}
\frac{1}{r} \gg T \gg \frac{\als}{r} \,.
\end{equation}
Interestingly, this case can be solved analytically.

Because $T$ is much larger than the typical energy scale,
the energy-dependent exponentials in \eqref{eq:sigma}, \eqref{eq:Xi}, \eqref{eq:sigmao}, \eqref{eq:Xi1} and \eqref{eq:Xi2} can be set equal to one.
We obtain
\begin{eqnarray}
(\Sigma_s +\Sigma_s^\dagger)\rho_s &=&  \Xi_{os} \,,
\\
(\Sigma_o +\Sigma_o^\dagger)\rho_o - \Xi_{oo}  &=&  \Xi_{so} = \frac{1}{N_c^2-1} (\Sigma_s +\Sigma_s^\dagger)\rho_o \,.
\end{eqnarray}
The evolution equations depend, therefore, on only one time-dependent parameter $\Gamma_s(t) = \Sigma_s(t) +\Sigma_s^\dagger(t)$,
the color-singlet width, and they read
\begin{eqnarray}
\frac{d\rho_{s\,r}(t;t)}{dt} = \Gamma_s(t) \left[\frac{\rho_{o\,r}(t;t)}{N_c^2-1}-\rho_{s\,r}(t;t)\right]\,,
\label{stat1}\\
\frac{d\rho_{o\,r}(t;t)}{dt} = - \Gamma_s(t) \left[\frac{\rho_{o\,r}(t;t)}{N_c^2-1}-\rho_{s\,r}(t;t)\right]\,.
\label{stat2}
\end{eqnarray}
The initial condition describes two heavy quarks at a given distance ${\bf r}$ in an arbitrary color state:
\begin{equation}
\rho_s(t_0;t_0)=\rho_{s\,r}(t_0;t_0)|{\bf r}\rangle\langle {\bf r}|\,,
\end{equation}
\begin{equation}
\rho_o(t_0;t_0)=\rho_{o\,r}(t_0;t_0)|{\bf r}\rangle\langle {\bf r}|\,.
\end{equation}
The problem consists in solving the evolution equations \eqref{stat1} and \eqref{stat2} for the two functions $\rho_{s\,r}$ and $\rho_{o\,r}$. 
The solution is
\begin{eqnarray}
\rho_{s\,r}(t;t) &=& \frac{\rho_{s\,r}(t_0;t_0)}{N_c^2} \left[1 + (N_c^2-1) e^{-\int_{t_0}^t dt'/u(t')}\right] 
+ \frac{\rho_{o\,r}(t_0;t_0)}{N_c^2}\left[1 - e^{-\int_{t_0}^t dt'/u(t')}\right]\,,
\label{eq:evorhosr} \\
\rho_{o\,r}(t;t) &=& \frac{\rho_{s\,r}(t_0;t_0)}{N_c^2}\,\left[N_c^2-1 - (N_c^2-1) e^{-\int_{t_0}^tdt'/u(t')}\right] 
+ \frac{\rho_{o\,r}(t_0;t_0)}{N_c^2}\left[N_c^2 - 1 + e^{-\int_{t_0}^t dt'/u(t')}\right],
\label{eq:evorhoor}
\end{eqnarray}
with $\displaystyle u(t) = \frac{N_c^2-1}{N_c^2}\frac{1}{\Gamma_s(t)}$.  
The static limit does not give us information on how the singlets are distributed in the different possible states 
(for example, $\Upsilon(1S)$, $\Upsilon(2S)$, and so on) but it does give qualitative information on how the population of singlets compares to that of octets. 
The crucial parameter is $u(t)$: for $t-t_0\ll u(t)$, the thermal medium has a small impact on the distribution of quarkonia,  
while for $t-t_0\gg u(t)$, the density approaches the large-time asymptotic value. \\

{\it (a)} As a first special situation, we consider a strongly-coupled plasma:
\begin{equation}
T \sim m_D \,.
\end{equation}
From \eqref{strong1} it follows that in this case $\Gamma_s(t) = \Sigma_s(t) +\Sigma_s^\dagger(t) = \kappa(t)\,r^2$.
The heavy-quark momentum diffusion coefficient $\kappa$ has been defined in \eqref{kappa}. 
The equations do not depend on the coefficient $\gamma$ defined in \eqref{gamma}.

\begin{figure}[ht]
\begin{center}
\includegraphics[scale=0.6]{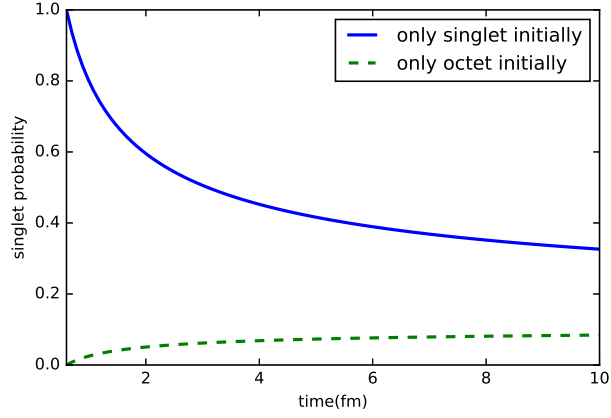}
\caption{Evolution in time of the color-singlet density for static quarks in a strongly-coupled plasma. 
The blue continuous line shows the evolution when the initial state is made only of singlets, 
whereas the green dashed line shows the evolution when the initial state is made only of octets.
Asymptotically both curves approach $1/N_c^2 = 1/9 \approx 0.11$.
\label{fig:staticstrong}
}
\end{center}
\end{figure}

We estimate the order of magnitude of $u(t)$ by taking $r=a_0$ (the Bohr radius of the $\Upsilon(1S)$), 
the average temperature $T=317$~MeV (that we define as the average between the temperature at $t=0.6$~fm and $T=250$~MeV) and $\kappa=2.5\,T^3$.
We obtain $u\sim 4$~fm, which is about the time the fireball temperature is above $T_c$ for central collisions (see Fig.~\ref{fig:Ttime}).
In Fig.~\ref{fig:staticstrong} we plot $\rho_{s}$ as a function of $t$ for the two extreme initial conditions.

{\it (b)} The second situation that we consider is the one of a weakly-coupled quark-gluon plasma: 
\begin{equation}
\frac{\als}{r} \gg m_D \,.
\end{equation}
From \eqref{weak1} in the infinite mass limit it follows that
$\displaystyle \Gamma_s(t) = \Sigma_s(t) +\Sigma_s^\dagger(t) = \frac{N_c(N_c^2-1)\,\als(\mu_E)\,\als^2(1/a_0)\,T}{6}$.

\begin{figure}[H]
\begin{center}
\includegraphics[scale=0.6]{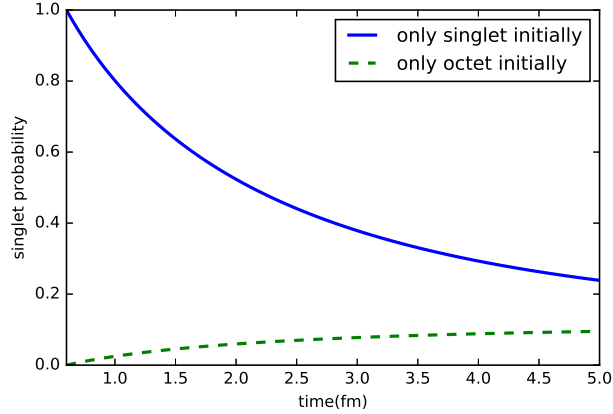}
\caption{Like in Fig.~\ref{fig:staticstrong}, but in a weakly-coupled quark-gluon plasma with the choice $\als \approx 0.4$ for all strong couplings.
\label{fig:staticweak}
}
\end{center}
\end{figure}

If we choose $\mu_E =  \pi T$, we get $u \approx 1.6$~fm. 
This is qualitatively consistent with what we see in Figs.~\ref{fig:E} and~\ref{E_running}. 
The computations done in~\cite{Brambilla:2010vq} were used in~\cite{Aarts:2011sm} to fit lattice results
using $\als \approx 0.4$ independently of the renormalization scale. 
If we proceed like that, we get $u \approx 2.16$~fm.
The plot of $\rho_{s}$ as a function of $t$ for the two extreme initial conditions
corresponding to this last choice of the coupling is shown in Fig.~\ref{fig:staticweak}. 

{\it (c)} Finally, let us consider the more general case where $\Gamma_s$ has an unspecified power-law dependence on the temperature: 
\begin{equation}
\Gamma_s(T)=\Gamma_s(T_0)\left(\frac{T}{T_0}\right)^n.
\label{GammapowerT}
\end{equation}
This case allows to understand, at least qualitatively, the interplay between the time evolution of the decay width and Bjorken's expansion
in the time evolution of the color-singlet and color-octet densities. 
Note that, if we neglect the running of the strong coupling, both previously considered situations, {\it (a)} and {\it (b)}, are of this type.

Using the time dependence of the temperature in Bjorken's expansion \eqref{Tbj}, one gets
\begin{equation}
  u(t)=\frac{N_c^2-1}{N_c^2}\frac{1}{\Gamma_s(T_0)}\left(\frac{t}{t_0}\right)^{nv_s^2}\,,
  \label{upower}
\end{equation}
which implies 
\begin{equation}
  \int_{t_0}^t\frac{\,dt'}{u(t')} = - \frac{N_c^2}{N_c^2-1}\;\frac{\Gamma_s(T_0)t_0}{1-nv_s^2}\left[ 1- \left(\frac{t}{t_0}\right)^{1-nv_s^2} \right].
\label{intu}
\end{equation}
Inserting \eqref{intu} into \eqref{eq:evorhosr} and \eqref{eq:evorhoor} provides the solution of the evolution equations:
\begin{equation}
  \rho_{s\,r}(t;t) = 1 - \rho_{o\,r}(t;t) = \frac{1}{N_c^2} 
  -\frac{N_c^2-1}{N_c^2}
  \left(\frac{\rho_{o\,r}(t_0;t_0)}{N_c^2-1}-\rho_{s\,r}(t_0;t_0)\right)e^{\frac{N_c^2}{N_c^2-1}\,\frac{\Gamma_s(T_0)t_0}{1-nv_s^2}\left[ 1- \left(\frac{t}{t_0}\right)^{1-nv_s^2} \right]}\,.
\label{eq:res}
\end{equation}

The solution shows three very different behaviors depending on the value of $nv_s^2$.
\begin{itemize}
\item[{\it (c.1)}]{If $nv_s^2>1$ the color-singlet density never reaches the value $1/N_c^2$,  which is its thermal equilibrium value.
  Instead it approaches the value
\begin{equation}
  \frac{1}{N_c^2}
   -\frac{N_c^2-1}{N_c^2}
  \left(\frac{\rho_{o\,r}(t_0;t_0)}{N_c^2-1}-\rho_{s\,r}(t_0;t_0)\right)e^{\frac{N_c^2}{N_c^2-1}\,\frac{\Gamma_s(T_0)t_0}{1-nv_s^2}}\,.
\end{equation}
The physical interpretation is that for $nv_s^2>1$, the decrease with time of the decay width in the fireball, described by Bjorken's expansion,
is so fast that the static quark-antiquark densities do not have time to equilibrate.}
\item[{\it (c.2)}]{In the case $nv_s^2<1$, we are exactly in the opposite situation.
  In the absence of freeze-out effects that could modify the evolution,
  the color-singlet and color-octet densities reach their thermal equilibrium values exponentially fast.
  This is the situation realized by static quarks and antiquarks in the weakly-coupled plasma of case {\it (b)}, for which $n=1$.
}
\item[{\it (c.3)}]{If $nv_s^2 = 1$,  we have
\begin{equation}
  \int_{t_0}^t\frac{\,dt'}{u(t')}=\frac{N_c^2}{N_c^2-1}\Gamma_s(T_0)t_0 \, \ln\left(\frac{t}{t_0}\right),
\end{equation}
which implies that 
\begin{equation}
 \rho_{s\,r}(t;t) = 1 - \rho_{o\,r}(t;t) = \frac{1}{N_c^2} 
  -\frac{N_c^2-1}{N_c^2}
  \left(\frac{\rho_{o\,r}(t_0;t_0)}{N_c^2-1}-\rho_{s\,r}(t_0;t_0)\right)
  \left(\frac{t_0}{t}\right)^{\frac{N_c^2}{N_c^2-1}\Gamma_s(T_0)t_0}\,.
\label{eq:resc3}
\end{equation}
Like before, the color-singlet and color-octet densities reach after some time their thermal equilibrium values.
Differently from the previous case, however, the falloff is power-like instead of exponential,
which means that the asymptotic values are reached slower (in fact much slower if $\Gamma_s(T_0)t_0 \ll 1$).
The physical interpretation is that we are describing a situation 
in which the speed of the expansion of the fireball is competing with the decrease of the decay width with the temperature.
The actual time required to reach the thermal equilibrium values depends crucially on the initial condition.
This situation is realized by static quarks and antiquarks in the strongly-coupled plasma of case {\it (a)}, for which $n=3$.
}
\end{itemize}

\section{Quarkonium in the regime $1/a_0 \gg T \gg E \gg m_D$}
\label{E>m_D}
In this appendix we consider quarkonium in a plasma that realizes the regime $1/a_0 \gg T \gg E \gg m_D$.
We aim at justifying Eqs. \eqref{eq:vs}-\eqref{weak5}.
We will start with some general remarks and then apply them to the case at hand.

\subsection{Density matrices' redefinitions}
The evolution equations \eqref{eq:ev} and \eqref{eq:ev_octet} allow for the following redefinitions of the density matrices $\rho_s$ and $\rho_o$ 
and of the functions $\Sigma_s$, $\Sigma_o$, $\Xi_{so}$, $\Xi_{os}$, and $\Xi_{oo}$ that preserve the evolution equations at the given accuracy
of order $r^2$ in the multipole expansion:
\bea
&&
\rho_s(t;t) \to \rho_s(t;t) - e^{-ih_s t}\,O_s(t)\,e^{ih_s t}\rho_s(t;t) - \rho_s(t;t) e^{-ih_s t}\,O_s(t)^\dagger\,e^{ih_s t} 
+ e^{-ih_s t}\,O_{so}(\rho_o(t;t),t)\,e^{ih_s t}, 
\label{redefinitionrhos}\\
&& \rho_o(t;t) \to \rho_o(t;t) - e^{-ih_o t}\,O_o(t)\,e^{ih_o t}\rho_o(t;t) - \rho_o(t;t) e^{-ih_o t}\,O_o(t)^\dagger\,e^{ih_o t} + e^{-ih_o t}\,O_{os}(\rho_s(t;t),t)\,e^{ih_o t} 
\nn\\
&& \hspace{10.8cm}
+ e^{-ih_o t}\,O_{oo}(\rho_o(t;t),t)\,e^{ih_o t},
\label{redefinitionrhoo}\\
\nn\\
&& \Sigma_s(t) \to \Sigma_s(t) + e^{-ih_s t}\,\frac{dO_s(t)}{dt}\,e^{ih_s t},
\label{redefinition1}\\
&& \Sigma_o(t) \to \Sigma_o(t) + e^{-ih_o t}\,\frac{dO_o(t)}{dt}\,e^{ih_o t},
\label{redefinition2}\\
&& \Xi_{so}(\rho_o(t;t),t) \to \Xi_{so}(\rho_o(t;t),t) + e^{-ih_s t}\,\frac{dO_{so}(\rho_o(t;t),t)}{dt}\,e^{ih_s t},
\label{redefinition3}\\
&& \Xi_{os}(\rho_s(t;t),t) \to \Xi_{os}(\rho_s(t;t),t) + e^{-ih_o t}\,\frac{dO_{os}(\rho_s(t;t),t)}{dt}\,e^{ih_o t},
\label{redefinition4}\\
&& \Xi_{oo}(\rho_o(t;t),t) \to \Xi_{oo}(\rho_o(t;t),t) + e^{-ih_o t}\,\frac{dO_{oo}(\rho_o(t;t),t)}{dt}\,e^{ih_o t},
\label{redefinition5}
\eea
where $O_s$, $O_o$, $O_{so}$, $O_{os}$ and $O_{oo}$ are operators of order $r^2$. 
This implies that at order $r^2$ we may neglect redefinitions of the density matrices inside the functions $\Xi_{so}$, $\Xi_{os}$ and $\Xi_{oo}$.
Moreover, whenever $d\rho_s/dt$ or $d\rho_o/dt$ multiplies functions of order $r^2$, we can replace them, through the leading-order evolution equations, 
with $-i[h_s,\rho_s]$ and $-i[h_o,\rho_o]$, respectively. 
In order for these transformations to preserve the trace of the total heavy-quark density, i.e., $\Tr \{\rho_s\} + \Tr \{\rho_o\}$, 
the operators $O_s$, $O_o$, $O_{so}$, $O_{os}$, and $O_{oo}$ must be related in such a way that the transformed 
color-singlet and color-octet density matrices, and the transformed functions $\Sigma_s$, $\Sigma_o$, $\Xi_{so}$, $\Xi_{os}$, and $\Xi_{oo}$ 
fulfill the conditions \eqref{conssin} and \eqref{consoct}. 
This is guaranteed if the final evolution equation is of the Lindblad form (see Sec.~\ref{sec:lindblad}).

\subsection{Computation of $\Sigma_s$ and $\Xi_{so}$}
Equations \eqref{eq:vs} and \eqref{weak1} were originally derived in~\cite{Brambilla:2010vq} by computing 
the singlet self energy in momentum space for a generic incoming energy and by dropping terms that would vanish on physical states. 
Such terms contribute to the wavefunction normalization only.
We may distinguish two momentum regions in the chromoelectric correlator appearing in the self energy.
The momentum region scaling like the temperature $T$ contributes with 
\bea
- i \frac{2\pi}{9}C_F\als T^2\,\left(-\Delta V r^2 - \frac{3}{M} \right),
\label{sp1}
\eea
where $\Delta V = h_o-h_s = \nc \als/(2r)$.
The momentum region scaling like the energy scale $E$ contributes with 
\be
\frac{2}{3}\als\,\cf T \left(\Delta V^2 r^2 +\frac{4 \Delta V }{M}+\frac{4\bp^2}{M^2}\right) .
\label{sp2}
\ee
The factor of $T$ appearing in \eqref{sp2} is a consequence of the Bose enhancement. 
The real part of $-i$ times \eqref{sp1} gives \eqref{eq:vs}, and the imaginary part of $-i$ times \eqref{sp2} gives \eqref{weak1}.

The function $\Sigma_s$ that appears in the evolution equation \eqref{eq:ev} has been defined in \eqref{eq:sigma}.
It also gets contributions from the scale $T$ and the scale $E$. 
The contribution from the scale $T$, which is computed by expanding the exponents and keeping terms linear in $h_o$ and $h_s$, reads 
\be
\left.\Sigma_s\right|_T = \hbox{Eq.~\eqref{sp1}}  + i \frac{\pi}{9} C_F \alpha_{\rm s}T^2 \left[ h_s , r^2\right].
\label{sr1}
\ee
The contribution from the scale $E$, which is computed by expanding the Bose distribution function for large $T$, reads
\be
\left.\Sigma_s\right|_E =  \hbox{Eq.~\eqref{sp2}} + \frac{2}{3} C_F \alpha_{\rm s}T \left[h_s, \frac{1}{2}\left[h_s, r^2\right] + r^2\Delta V \right].
\label{sr4}
\ee

We see that neither \eqref{sr1} nor \eqref{sr4} coincides with \eqref{sp1} and \eqref{sp2}. 
The difference is significant: for instance, in \eqref{sr1} it gives rise to an imaginary part of $-i\Sigma_s$ that is larger than the one listed in \eqref{weak1}.
However, assuming that the time-dependence of the temperature is a subleading effect (see the comment at the end of Sec.~\ref{sec:evolution}), we may reabsorb the difference 
in a redefinition of the color-singlet density, $\rho_s$, and $\Sigma_s$ according to \eqref{redefinition1} with
\be
O_s(t) = e^{ih_s t}\left\{  \frac{\pi}{9} C_F \alpha_{\rm s}T^2 r^2 
- i \frac{2}{3} C_F \alpha_{\rm s}T  \left( \frac{1}{2}\left[h_s, r^2\right] + r^2\Delta V \right) \right\}e^{-ih_s t}.
\label{Os}
\ee

Also $\Xi_{so}$, defined in \eqref{eq:Xi}, may be computed in a similar way by distinguishing contributions coming from the momentum region $T$,
\be
\left.\Xi_{so}\right|_T  = i \frac{\pi}{9} \frac{\alpha_{\rm s}}{\nc} T^2  \left( r^i \left[ \rho_o, h_o \right] r^i -\left[ r^i\rho_o r^i, h_s \right] \right) ,
\label{XisoT}
\ee
from contributions coming from the momentum region $E$,
\be
\left.\Xi_{so}\right|_E  = \hbox{Eq.~\eqref{weak3}}  
+ \frac{\alpha_{\rm s}}{3\nc} T  \left( r^i\left[\left[\rho_o\,, h_o\right], h_o\right] r^i- 2 \left[r^i\left[ \rho_o, h_o\right] r^i, h_s\right] +\left[\left[r^i\rho_o r^i, h_s \right], h_s\right]\right) .
\label{XisoE}
\ee
We see again that the sum of \eqref{XisoT} and \eqref{XisoE} differs from \eqref{weak3} by a quantity that may be reabsorbed 
into a redefinition of the color-singlet density, $\rho_s$, and $\Xi_{so}$ according to \eqref{redefinition3} with
\be
O_{so}(\rho_o(t;t),t) = \frac{\pi}{9} \frac{\alpha_{\rm s}}{\nc} T^2 e^{ih_s t}  r^i \rho_o r^i  e^{-ih_s t}
- \frac{\alpha_{\rm s}}{3\nc} T \frac{d}{dt}\left(    e^{ih_s t} r^i \rho_o r^i   e^{-ih_s t}\right).
\label{Oso}
\ee
As mentioned above, the redefinition relies on the time dependence of the temperature being a subleading effect and on using the leading-order evolution equations 
whenever this is consistent with the order $r^2$ accuracy of the calculation.

\subsection{Computation of $\Sigma_o$, $\Xi_{os}$, and $\Xi_{oo}$}
\label{octet}
The octet thermal corrections \eqref{eq:vo} and \eqref{weak2}, which are related to the octet potential and width, have been written here for the first time.
They follow from the results in~\cite{Brambilla:2010vq} in a straightforward way.
Two diagrams contribute to the octet self energy, one with a singlet propagator in the loop and one with an octet propagator in it, 
because, according to the pNRQCD Lagrangian \eqref{eq:Lagrangian_pNRQCD}, the octet may couple to both the singlet and the octet itself through a chromoelectric dipole.  
The contributions of the first diagram can be obtained from the ones of the singlet self energy \eqref{sp1} and \eqref{sp2} by making the substitution 
$h_o \leftrightarrow h_s$ ($\Delta V \to -\Delta V$) and correcting the global color factor $C_F \to 1/(2N_c)$. 
This (times~$-i$) leads to the first two and three terms in \eqref{eq:vo} and \eqref{weak2}, respectively.
The contributions of the second diagram can be obtained in a similar way by making the substitutions $h_s \to h_o$ ($\Delta V=0$) and $C_F \to (\nc^2-4)/4\nc$.
This (times~$-i$) leads to the last term in both \eqref{eq:vo} and \eqref{weak2}.

The computation of $\Sigma_o$ from \eqref{eq:sigmao} leads to an expression that differs from \eqref{eq:vo} and \eqref{weak2} by an amount 
that can be reabsorbed into a redefinition of the color-octet density, $\rho_o$, and $\Sigma_o$ according to \eqref{redefinition2} with
\bea
O_o(t) &=& e^{ih_o t}\left\{  \frac{\pi}{18} \frac{\alpha_{\rm s}}{\nc} T^2 r^2 
- \frac{i}{3} \frac{\alpha_{\rm s}}{\nc} T  \left( \frac{1}{2}\left[h_o, r^2\right] - r^2\Delta V \right) 
\right.
\nn\\
&& \hspace{10mm}
\left.+\frac{\pi}{36} \frac{\nc^2-4}{\nc} \alpha_{\rm s}T^2 r^2 - \frac{i}{12} \frac{\nc^2-4}{\nc} \alpha_{\rm s}T \left[h_o, r^2\right] \right\}e^{-ih_o t}.
\label{Oo}
\eea

The functions $\Xi_{os}$ and $\Xi_{oo}$ are computed from \eqref{eq:Xi1} and \eqref{eq:Xi2}.
The calculation proceeds like in the $\Xi_{so}$ case discussed above, and the results may be copied from there after having performed the same substitutions 
that allow us to compute the two diagrams of $\Sigma_o$ from the one contributing to $\Sigma_s$.
The results one obtains differ from the expressions listed in \eqref{weak4} and \eqref{weak5} by an amount that can be reabsorbed into a redefinition of the color-octet density, $\rho_o$, 
and  $\Xi_{os}$ and $\Xi_{oo}$ according to \eqref{redefinition4} and \eqref{redefinition5} with
\bea
O_{os}(\rho_s(t;t),t) &=& 
\frac{\pi}{9} \frac{\alpha_{\rm s}}{\nc} (\nc^2-1) T^2 e^{ih_o t}  r^i \rho_s r^i  e^{-ih_o t}
- \frac{\alpha_{\rm s}}{3\nc} (\nc^2-1) T \frac{d}{dt}\left(    e^{ih_o t} r^i \rho_s r^i   e^{-ih_o t}\right),
\label{Oos}\\
O_{oo}(\rho_o(t;t),t) &=& 
\frac{\pi}{18} \frac{\alpha_{\rm s}}{\nc} (\nc^2-4) T^2 e^{ih_o t}  r^i \rho_o r^i  e^{-ih_o t}
- \frac{\alpha_{\rm s}}{6\nc} (\nc^2-4) T \frac{d}{dt}\left(    e^{ih_o t} r^i \rho_o r^i   e^{-ih_o t}\right).
\label{Ooo}
\eea
Finally, we note that $\Tr \{\rho_s\} + \Tr \{\rho_o\}$ is invariant under the density redefinitions induced by the functions \eqref{Os} and \eqref{Oso}-\eqref{Ooo}.

\subsection{Impact on $R_{AA}(nS)$}
According to \eqref{eq:raa} and \eqref{rhosdef} the nuclear modification factor, $R_{AA}(nS)$, for $S$-wave quarkonia is proportional to the matrix element
\be
\langle nS| \rho_s(t;t)|nS\rangle\,.
\ee
A redefinition of the color-singlet density matrix of the type \eqref{redefinitionrhos} changes the above matrix element into
\bea
\langle nS| \rho_s(t;t)|nS\rangle &\to& 
\langle nS| \rho_s(t;t)|nS\rangle - \langle nS|e^{-ih_st}O_s(t)e^{ih_st}\rho_s(t;t)|nS\rangle - \langle nS|\rho_s(t;t)e^{-ih_st}O_s(t)^\dagger e^{ih_st}|nS\rangle 
\nn\\
&& 
\hspace{2.5cm}
+ \langle nS| O_{so}(\rho_o(t;t),t) | nS \rangle\,.
\label{redefinitionmatrix}
\eea
This shift in the color-singlet matrix element has been taken into account when computing the results shown in Sec.~\ref{sec:weak:results}.
Its numerical impact turns out to be small.

\section{Quarkonium in the regime $1/a_0 \gg T \gg m_D \gg E$}
\label{m_D>E}
We consider here quarkonium in a plasma that realizes the regime $1/a_0 \gg T \gg m_D \gg E$.
This regime was studied in~\cite{Brambilla:2008cx} for the static case and in~\cite{Escobedo:2010tu} for muonic hydrogen.
It is straightforward to combine those results to obtain the relevant expansions that enter the Lindblad equation.
Under the condition $1/a_0 \gg T \gg m_D \gg E$, all thermal contributions can be encoded in modifications of the potential. 
In addition, most of the contributions can be regarded as a particular case of Sec.~\ref{sec:strong},
where $\gamma$ and $\kappa$ can be computed in perturbation theory.
The expected leading contributions to $\gamma$ and $\kappa$, which would be of ${\cal O}(\als (\mu_T) T^3)$, turn out to vanish.
As a consequence, the leading non-vanishing contributions to these quantities are of ${\cal O}(\als^2 (\mu_T) T^3)$ and ${\cal O}(\als (m_D) m_D^2 T)$.
They may compete in size with terms of relative size of ${\cal O}(\als (\mu_T) E T^2)$ calculated in \eqref{eq:vs}
that come from higher-order terms in the $E/T$ expansion and are not included in $\gamma$ and $\kappa$.
Putting all the contributions together, we obtain,
\begin{eqnarray}
&& {\rm Re}\,(-i\Sigma_s)=\frac{2\pi}{9} C_F\alpha_{\rm s}(\mu_T) \Delta V T^2r^2+\frac{2\pi}{3M}C_F\alpha_{\rm s}(\mu_T)T^2+\frac{r^2}{2}\gamma \,,
  \label{appC1Sigmas}
  \\
&& \gamma=-3\zeta(3)C_F\frac{\alpha_{\rm s}(m_D)}{\pi}Tm_D^2+\frac{4}{3}\zeta(3)N_c C_F\alpha_{\rm s}^2(\mu_T)T^3 \,,
\label{gammapert}
\\
\nn
\\
&& {\rm Im}\,(-i\Sigma_s)=-\frac{r^2}{2}\kappa\,,
\\
&& \kappa=-\frac{C_F}{3}\alpha_{\rm s}(m_D) Tm_D^2\left(2\gamma_E-\log\frac{T^2}{m_D^2}-1-4\log 2-2\frac{\zeta'(2)}{\zeta(2)}\right)-\frac{8\pi}{9}\log 2N_cC_F\alpha_{\rm s}^2(\mu_T)T^3\,,
\label{eq:vsp}
\end{eqnarray}
where $\mu_T=\pi T$, $m_D^2= 4\pi\alpha_{\rm s}(\mu_T)T^2(N_c+n_f/2)/3$, and $\Delta V= N_c \alpha_{\rm s}(1/a_0)/(2r)$;
$n_f$ is the number of active massless flavors ($n_f=3$ in the bottomonium case); and $\zeta$ is the Riemann zeta function. 
Note that the term $r^2\gamma/2$ in \eqref{appC1Sigmas} is suppressed by a factor $a_0T$ with respect to the other terms;
we keep it nevertheless to maintain a closer analogy to the discussion of Sec.~\ref{sec:strong}.

The thermal contributions to the octet potential have not been calculated before.
However, they can be easily obtained from the expressions above by following the same steps as in Appendix~\ref{octet}. We obtain
\begin{eqnarray}
  &&{\rm Re}\,(-i\Sigma_o)=-\frac{\pi}{9 N_c}\alpha_{\rm s}(\mu_T)\Delta V T^2r^2
  + \frac{\pi}{6M}\frac{N_c^2-2}{N_c}
  \alpha_{\rm s}(\mu_T)T^2
  +\frac{r^2}{4}
  \frac{N_c^2-2}{N_c^2-1}
  \gamma\,,
  \\
  && {\rm Im}\,(-i\Sigma_o)= -\frac{r^2}{4}
  \frac{N_c^2-2}{N_c^2-1}
  \kappa  \,.
\end{eqnarray}
The quantities $\Xi_{so}$, $\Xi_{os}$, and $\Xi_{oo}$ are obtained as particular cases of \eqref{strong3}-\eqref{strong5}, 
in which $\gamma$ and $\kappa$ take the values of \eqref{gammapert} and \eqref{eq:vsp}, respectively.
In addition, the Lindblad operator $H$ is obtained from \eqref{Hlindstrong} by making the following replacements,
\bea
h_s & \to & h_s + \frac{2\pi}{9} C_F\alpha_{\rm s}(\mu_T) \Delta V T^2r^2 + \frac{2\pi}{3M}C_F\alpha_{\rm s}(\mu_T)T^2\,, 
\\
h_o & \to & h_o - \frac{\pi}{9 N_c} \alpha_{\rm s}(\mu_T) \Delta V T^2r^2 + \frac{\pi}{6M}\frac{N_c^2-2}{N_c}\alpha_{\rm s}(\mu_T)T^2\,.
\eea

\begin{figure}
\includegraphics[scale=0.5]{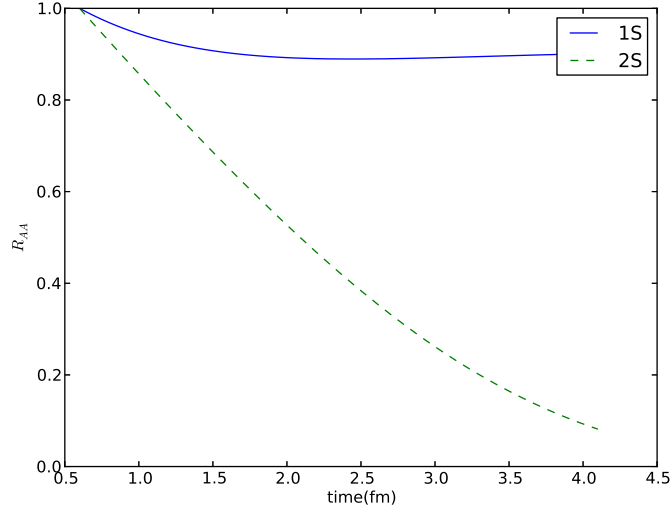}
\caption{$R_{AA}$ for $\Upsilon(1S)$ and $\Upsilon(2S)$ due to screening only, in the regime $1/a_0 \gg T \gg m_D \gg E$.}
\label{fig:screening}
\end{figure}

It is not obvious that this regime applies to LHC temperatures. In particular the separation of scales between $T$ and $m_D$ does not seem to be
large enough to guarantee a positive $\kappa$ and hence a positive decay width.
We can nevertheless compute the quarkonium suppression just due to screening, which amounts to computing 
$R_{AA}$ while ignoring the imaginary part of the potential and all the $\Xi$'s.
In Fig.~\ref{fig:screening}, this suppression is computed from \eqref{eq:raa} in the case of the $1S$ and $2S$ bottomonium states.
We can see that, while screening alone is able to make $\Upsilon(2S)$ disappear almost completely, $\Upsilon(1S)$ suppression is at most 10\%.

\section{Alternative form of the evolution equations}
\label{linear}
We present here an alternative way of writing the selfenergies and the $\Xi$ functions in the evolution equations (\ref{eq:ev})-(\ref{eq:ev_octet}),
which makes the linear character of the equations and the minimal set of operators that characterize the interactions with the medium explicit:
\begin{eqnarray}
&& \Sigma_s(t) = r^i \, A_i^{so\,\dagger}(t), \nn\\
&& \Sigma_o(t) = \frac{1}{N_c^2-1} r^i \,A_i^{os\,\dagger}(t) + \frac{N_c^2-4}{2(N_c^2-1)} r^i \,A_i^{oo\,\dagger}(t), \nn\\
&& \Xi_{so}(\rho_o(t;t),t) = \frac{1}{N_c^2-1}\left(A_i^{os\,\dagger}(t) \, \rho_o(t;t) \, r^i  + r^i \, \rho_o(t;t) \, A_i^{os}(t)   \right),\label{lineareqs}\\
&& \Xi_{os}(\rho_s(t;t),t) = A_i^{so\,\dagger}(t)  \rho_s(t;t)  r^i + r^i \rho_s(t;t)A_i^{so}(t) ,\nn\\
&& \Xi_{oo}(\rho_o(t;t),t) = \frac{N_c^2-4}{2(N_c^2-1)}\left(A_i^{oo\,\dagger}(t) \, \rho_o(t;t) \,  r^i  + r^i \, \rho_o(t;t) \, A_i^{oo}(t)   \right),\nn
\end{eqnarray}
where
\be
A_i^{uv}(t) =  \frac{g^2}{2N_c}\int_{t_0}^{t} dt_2 \, e^{ih_u (t_2-t)} \, r^j \, e^{ih_v (t-t_2)} \, \langle E^{a,j}(t_2,{\bf 0})E^{a,i}(t,{\bf 0})\rangle \,,
\ee
with $u,v=s,o$.

We can use Eqs. \eqref{lineareqs} to write the evolution equation in a way similar to the Lindblad equation. 
Defining 
\begin{equation}
\rho=\left(\begin{array}{cc}
\rho_s & 0\\
0 & \rho_o\end{array}\right)\,,\hspace{2cm}H=\left(\begin{array}{cc}
h_s+\frac{\Sigma_s-\Sigma_s^\dagger}{2i} & 0\\
0 & h_o+\frac{\Sigma_o-\Sigma_o^\dagger}{2i}
\end{array}\right)\,,
\end{equation}
\begin{equation}
L_i^0=\left(\begin{array}{cc}
0 & 0 \\
0 & 1
\end{array}\right)\, r^i\,,\hspace{2cm}L_i^1=\left(\begin{array}{cc}
0 & 0 \\
0 & \frac{N_c^2-4}{2(N_c^2-1)}{A_i^{oo}}^\dagger
\end{array}\right)\,,
\end{equation}
\begin{equation}
L_i^2=\left(\begin{array}{cc}
0 & 1 \\
1 & 0
\end{array}\right)\, r^i\,,\hspace{2cm}L_i^3=\left(\begin{array}{cc}
0 & \frac{1}{N_c^2-1}{A_i^{os}}^\dagger \\
{A_i^{so}}^\dagger & 0
\end{array}\right)\,,
\end{equation}
the evolution equation can be written as 
\begin{equation}
\frac{d\rho}{dt}=-i[H,\rho]+\sum_{nm}h_{nm}\left(L_i^n\rho {L_i^m}^\dagger-\frac{1}{2}\{{L_i^m}^\dagger L_i^n,\rho\}\right)\,,
\end{equation}
where $h_{nm}$ are the elements of the matrix 
\begin{equation}
h=\left(\begin{array}{cccc}
0 & 1 & 0 & 0 \\
1 & 0 & 0 & 0 \\
0 & 0 & 0 & 1 \\
0 & 0 & 1 & 0 
\end{array}\right)\,.
\end{equation}
If $h$ were a positive definite matrix, then it would always be possible to redefine the operators $L_i^n$ in such a way 
that the evolution equation would be of the Lindblad form \eqref{eq:Lindblad}. 
Since, however, $h$ is not a positive definite matrix, the Lindblad theorem~\cite{Lindblad:1975ef} does not guarantee that 
Eqs. \eqref{eq:ev} and \eqref{eq:ev_octet} may be brought into a Lindblad form.
A special case is the strongly-coupled case studied in Sec.~\ref{sec:strong}.
There $L_i^1\propto L_i^0$ and $L_i^3\propto L_i^2$, which allows us to set to zero, after a redefinition of the operators $L_i^n$, the eigenvectors of $h$ associated to negative eigenvalues, 
eventually leading to an evolution equation of the Lindblad form.

\bibliography{fireball}
\bibliographystyle{apsrev4-1}

\end{document}